\documentclass[twocolumn,aps,pra,showpacs,raggedbottom,nobalancelastpage,amssymb,superscriptaddress,longbibliography,nofootinbib]{revtex4-2}
\usepackage{graphicx}
\usepackage{amsmath}
\usepackage{amssymb}
\usepackage{physics}
\usepackage{bm}
\usepackage[usenames]{color}
\usepackage{amsfonts}
\usepackage[centering,hmargin=2cm,vmargin=2.5cm]{geometry}
\usepackage[dvipsnames]{xcolor}

\usepackage{comment}

\usepackage[colorlinks=true , citecolor=blue,urlcolor=blue]{hyperref}

\def\kbar{{\mathchar'26\mkern-9mu k}}

\usepackage{babel}
\usepackage{ulem}
\begin{document}
	\title[Article Title]{Localization properties of the asymptotic density distribution of a one-dimensional disordered system}
	\date{version 0, \today}
	
	\author{Cl\'ement Hainaut}
	\affiliation{Universit\'e de Lille, CNRS, UMR 8523 \textendash{} PhLAM \textendash{} Laboratoire de Physique des Lasers Atomes et Mol\'ecules, F-59000 Lille, France}
	\author{Jean-Fran\c cois Cl\'ement}
	\affiliation{Universit\'e de Lille, CNRS, UMR 8523 \textendash{} PhLAM \textendash{} Laboratoire de Physique des Lasers Atomes et Mol\'ecules, F-59000 Lille, France}
	\author{Pascal Szriftgiser}
	\affiliation{Universit\'e de Lille, CNRS, UMR 8523 \textendash{} PhLAM \textendash{} Laboratoire de Physique des Lasers Atomes et Mol\'ecules, F-59000 Lille, France}
	\author{Jean Claude Garreau}
	\affiliation{Universit\'e de Lille, CNRS, UMR 8523 \textendash{} PhLAM \textendash{} Laboratoire de Physique des Lasers Atomes et Mol\'ecules, F-59000 Lille, France}
	\author{Adam Ran\c con}
	\affiliation{Universit\'e de Lille, CNRS, UMR 8523 \textendash{} PhLAM \textendash{} Laboratoire de Physique des Lasers Atomes et Mol\'ecules, F-59000 Lille, France}
	\author{Radu Chicireanu}
	\affiliation{Universit\'e de Lille, CNRS, UMR 8523 \textendash{} PhLAM \textendash{} Laboratoire de Physique des Lasers Atomes et Mol\'ecules, F-59000 Lille, France}
		
\begin{abstract}
	Anderson localization is the ubiquitous phenomenon of inhibition of transport of classical and quantum waves in a disordered medium. In dimension one, it is well known that all states are localized, implying that the distribution of an initially narrow  wave-packet released in a disordered potential will, at long time, decay exponentially on the scale of the localization length. However, the exact shape of the stationary localized distribution differs from a purely exponential profile and has been computed almost fifty years ago by Gogolin.
	
	Using the atomic quantum kicked rotor, a paradigmatic quantum simulator of Anderson localization physics, we study this asymptotic distribution by two complementary approaches. First, we discuss the connection of the statistical properties of the system's localized eigenfunctions and their exponential decay with the localization length of the Gogolin distribution.  Next, we make use of our experimental platform, realizing an ideal Floquet disordered system, to measure the long-time probability distribution and highlight the very good agreement with the analytical prediction compared to the purely exponential one over 3 orders of magnitude.
\end{abstract}
	
\maketitle

\section{Introduction}
Anderson localization \cite{Anderson:LocAnderson:PR58,Abrahams:Scaling:PRL79}, the complete absence of transport due to disorder-induced destructive quantum interference, has been predicted more than 60 years ago and triggered enormous inspiration in the field of both classical and quantum transport. The main distinctive property of Anderson localization is the specific exponentially localized form of its wave-packet and has been observed in various physical systems ranging from light waves \cite{Wiersma:LightLoc:N97,Schwartz:LocAnderson2DLight:N07}, microwaves \cite{Dalichaouch1991,Chabanov2000}, sound waves \cite{WEAVER1990}, electrons gases \cite{AkkermansMontambaux:MesoscopicPhysics:11} and atomic matter waves \cite{Billy:AndersonBEC1D:N08,Chabe:Anderson:PRL08}, making it an ubiquitous feature in physics of wave transport in disordered media.

It is well known that in one dimension, all eigenstates of a disordered Hamiltonian decay exponentially \cite{Abrahams:Scaling:PRL79}. This implies that an initially narrow wave-packet will start to expand in the disordered medium until reaching a localized steady-state. The characteristics of this localized state are given by the probability to find a particle at a distance $x$ from its initial position which is given by the so-called Gogolin distribution \cite{Gogolin1976},
\begin{equation}
	\Pi_G(x)=\frac{\pi^2}{16\xi }\int z\sinh(\pi z)\left(\frac{1+z^2}{1+\cosh(\pi z)}\right)^2 e^{-\frac{1+z^2}{4\xi}\vert x \vert} \dd z.
	\label{eq:Gogolin}
\end{equation}
This distribution depends only on one parameter, the localization length $\xi$. In this sense, the shape of the distribution is `universal', whereas the parameter $\xi$, which describes the localization properties of the system, depends on the microscopic details, such as the disorder strength and energy of the initial state.

While this distribution has first been computed for a strictly one-dimensional system using Berezinskii diagrammatic method~\cite{Gogolin1975}, it was soon realized that it applies to a much wider class of systems. In particular, it also describes the asymptotic behavior of a broad class of systems, described by Efetov's supersymmetric non-linear sigma model 
\cite{Efetov:SupersymmetryInDisorder:97}, such as quasi-one-dimensional disordered systems \cite{Efetov1983a}, random band matrices \cite{Fyodorov1991}, and the Quantum Kicked Rotor (QKR) \cite{Altland1996}. The latter is a paradigmatic model of quantum chaos \cite{Izrailev:LocDyn:PREP90}, where a quantum particle is kicked periodically with a sinusoidal potential, and which displays dynamical localization \cite{Casati:LocDynFirst:LNP79}: an initially narrow wave-packet in \textit{momentum space} will reach at long time an exponentially localized  \textit{momentum distribution}. The connection between dynamical localization and Anderson localization has been explicitly realized by mapping the Kicked Rotor problem on a disordered tight-binding Hamiltonian, similar to the Anderson model~\cite{Fishman:LocDynAnders:PRL82}. Thus, the Floquet eigenstates of the evolution operator are exponentially localized, with a \textit{similar} localization length for all eigenstates.

This makes the experimental realization of the QKR a remarkable quantum simulator of the physics of disordered systems \cite{Garreau2017}. Indeed, the atomic QKR has allowed the observation of dynamical localization \cite{Moore:AtomOpticsRealizationQKR:PRL95,Moore:LDynFirst:PRL94}, Anderson localization in two dimensions \cite{Manai:Anderson2DKR:PRL15}, the Anderson transition in three dimensions \cite{Chabe:Anderson:PRL08,Lopez:ExperimentalTestOfUniversality:PRL12} detailed studies of weak-localization effects~\cite{Hainaut:ERO:PRL17,Hainaut2018CFS,Hainaut2018_Ratchet,Hainaut2018}

The creation of narrow and monochromatic matter wavepackets is generally very difficult experimentally, which is expected to hinder a precise observation of the Gogolin density profile. Indeed, the energy dependence of the localization length in `usual' disordered systems makes the asymptotic localized state a superposition of such profiles. The energy-independence of the QKR Floquet states is thus a very important asset in this respect, both numerically and experimentally.

In this paper, we investigate the asymptotic localization properties of an Anderson-localized wavepacket in the atomic QKR. Using numerical simulations, we show that the length $\xi$ that appears in the Gogolin distribution corresponds to the average localization length of the Floquet eigenstates, the distribution of which is shown to be in good agreement with the supersymmetric predictions for the eigenstate statistics. Then, using our experimental platform of the atomic QKR, we show that the long-time momentum distribution  is in excellent agreement with the Gogolin distribution (convolved with the initial momentum distribution), while a purely exponential form does not fit the data.

Our manuscript is organized as follows. In Sec.~\ref{sec_th}, we recall the basic properties of the QKR as well as the relevant theoretical predictions, and present some numerical simulations, in very good agreement with the supersymmetric predictions. We present our experimental observation of the Gogolin distribution in Sec.~\ref{sec_exp}, and present our conclusions in Sec.~\ref{sec_concl}.

\section{Statistical properties of the QKR eigenstates \label{sec_th}}

The QKR Hamiltonian is defined as
\begin{equation}
	\hat H_{\textrm{QKR}}=\frac{\hat p^{2}}{2}+K\sum_{n=0}^{\infty}\cos \hat x\;\delta\left(t-n\right),\label{eq:QKR}
\end{equation}
where $K\cos \hat x$ represents a sinusoidal potential created by a standing wave (formed by counterpropagating lasers of wave number $k_{L}$), with length in units of $\left(2k_{L}\right)^{-1}$ and time in units of the kick period $T_{1}$. Momenta are measured in units such that $\hat x$ and $\hat  p$ obey the canonical commutation relation $\left[\hat x,\hat p\right]=i\kbar$ with an effective Planck constant $\kbar=4\hbar k_{L}^{2}T_{1}/M$ (for particles of mass $M$).
The kick strength $K$, as well as $\kbar$, can be tuned in the experiment (see below). 
For an initial state with a well defined initial momentum, one observes that the kinetic energy of the system initially grows linearly in time, before saturating to a constant value, the hallmark of dynamical localization.

Due to the spatial periodicity of the potential, the kicks can only change the momentum by increments of $\kbar$, and writing momenta $p=(q+\ell)\kbar$, with $\ell\in\mathbb{Z}$ and $q\in(-1/2,1/2]$, the quasi-momentum $q$ is a conserved quantity. The evolution operator over one period reads, for a given quasi-momentum $q$\footnote{The dependence on quasi-momentum of all quantities is left implicit from now on.}:
\begin{equation}
	\hat U (1)=\exp\left(-i\frac{(\hat{\ell}+q)^{2}}{2}\kbar\right)\exp\left(-i\frac{K}{\kbar}\cos\hat{x}\right),
	\label{eq:U(1)}
\end{equation}
with $\hat{\ell}\left\vert \ell\right\rangle =\ell\left\vert \ell\right\rangle$, and the operator splits into a kicking part and a free propagation due to the instantaneous character of the kicks\footnote{In practice, it suffices that the kick duration is short enough that $p T_1/M\ll\lambda_L$, $\lambda_L=2\pi/k_L$.}.
Dynamical localization can be understood by noting that the Floquet eigenstates $\hat U\left\vert \phi_{\omega}\right\rangle =\exp(-i\omega)\left\vert \phi_{\omega}\right\rangle $ are, up to some technicalities, eigenstates of a disordered tight-binding Hamiltonian $\hat H_{\rm eff}$ displaying Anderson localization ~\cite{Fishman:LocDynAnders:PRL82,Fishman:LocDynAnderson:PRA84,Shepelyansk:KRFloquet:PRL86}. For the QKR, one finds
\begin{equation}
	\hat H_{\rm eff}=\sum_\ell\epsilon_{\ell}\left\vert \ell\right\rangle \left\langle \ell\right\vert +\sum_{\ell,\ell'}t_{\vert\ell-\ell'\vert}\left\vert \ell\right\rangle \left\langle \ell'\right\vert ,\label{eq:HA}
\end{equation}
with on-site energy $ \epsilon_{\ell}=\tan\left(\omega/2-\kbar{(\ell+q)}^{2}/4\right)$
and   hopping amplitude $t_{r}=(2\pi)^{-1}\int_0^{2\pi} \dd xe^{-irx}\tan\left(K\cos x/2\kbar\right)$. The on-site energies are deterministic, but for $\kbar$ incommensurate with $\pi$, they oscillate strongly enough to play the role of a pseudo-disorder, while each $q$ plays the role of a different disorder realization. Finally, the hopping $t_{r}$ has a range of order $K/\kbar$, and the Hamiltonian is thus similar to a random band matrix in the limit $K/\kbar\gg 1$.

It comes out of the mapping of Ref.\,~\cite{Fishman:LocDynAnders:PRL82} that all Floquet eigenstates are eigenvectors of $\hat H_{\rm eff}$ with zero energy, and are thus expected to have the same localization properties (e.g. same localization length). This is in contrast with disordered systems and random band matrices, where the localization properties depend on the position of the states in the spectrum. 

\bigskip

The QKR can be described by the same supersymmetric field theory that is used to describe random band matrices and quasi one-dimensional systems, and as a consequence possesses the same universal features. The characterization of these features can be done by studying the statistical properties of the system's spectrum and eigenstates, as was initially recognized by Wigner \cite{Wigner1955}, starting the field of Random Matrix Theory \cite{MehtaBook2}. In the present context, the supersymmetric method has allowed for detailed calculation of the statistics of the eigenstates, see \cite{Mirlin2000} for a review. Following~\cite{Fyodorov1993a,Mirlin2000}, we recall here only the relevant results necessary for the discussion.

The exponential decay of an eigenstate $\vert \phi_\omega\rangle$, $\vert \phi_\omega(\ell)\vert ^2\propto e^{-\vert \ell\vert/\xi_\omega}$  can be characterized by introducing the quantity $r_\omega(\ell,L)=\vert\phi_\omega(\ell)\vert^2\vert\phi_\omega(\ell+L)\vert^2$, since 
\begin{equation}
	\xi_\omega^{-1}=-\lim_{L\to \infty}L^{-1}\ln r_\omega(\ell,L).\label{eq:xiOmega}
\end{equation}
The statistical properties of $r_\omega$ have been studied in detail \cite{Mirlin2000}, as it also describes the Inverse Participation Ratio (IPR) of each eigenstate $P_\omega=\sum_\ell r_\omega(\ell,0)$, as well as the asymptotic long-time probability $\Pi(\ell,L)$ to find a particle in $\ell+L$ knowing that it has started at $\ell$, for a given disorder realization, 
\begin{equation}
	\Pi(\ell,L)=\sum_\omega r_\omega(\ell,L).
\end{equation}

Denoting with a bracket  the average over disorder realisations, for a narrow energy part of the spectrum, the probability distribution of $v=-\ln r_\omega$ has been shown to be Gaussian in the limit $L\gg \tilde\xi$ \cite{Fyodorov1993a}, with  $\tilde\xi^{-1}=\langle v\rangle/L$ the average inverse localization length of the eigenstates,
\begin{equation}\label{eq:distribPofX}
	\mathcal{P}(v)=\frac{\exp\left(-\frac{(v-\langle v\rangle)^2}{4\langle v\rangle}\right)}{\sqrt{4\pi\langle v\rangle}}.
\end{equation}
Noting that the variance of $v$ is twice its mean, this implies that the distribution of $1/\xi_\omega$ is sharply peaked as $L\to\infty$, and therefore $\langle \xi_\omega\rangle=\tilde\xi$.

Using similar supersymmetric methods, one shows that the momentum distribution at long time is given by the Gogolin distribution\footnote{The result is independent from the starting position $\ell$ thanks to the translation invariance after averaging over disorder.}
\begin{equation}
	\langle\Pi(\ell,L)\rangle =\Pi_G(L),
\end{equation}
with the same localization length as the Floquet eigenstates, i.e. $\xi=\tilde\xi$~\cite{Efetov1983a,Fyodorov1993a}.

Lastly, we point out that, as in the QRK the all the Floquet eigenstates have the same eigenenergy, the energy selection introduced in~\cite{Fyodorov1993a} is no longer required, and can be replaced (and used in the following section) by a broader averaging\footnote{Also denoted with brackets from now on, for consistency.}, over both disorder and the whole ensemble of eigenstates.

\section{Numerical investigations} \label{sec_num}
We shall now investigate numerically the relation between the typical localization length of the Kicked Rotor's Floquet eigenstates and the Gogolin distribution (see also~\cite{Casati1991,Dittrich1991,Izrailev1995} for early numerical studies of the spectral properties of the QKR).  In order to realize an accurate investigation, we will make use of an idealized version of the model, the so-called Random Kicked Rotor (RKR) where the kinetic term $\frac{(\hat{\ell}+q)^{2}}{2}\kbar$ is replaced by a purely random, uniformly distributed phase $\theta_\ell\in[0,2\pi[$. This allows us to suppress the undesired correlation effects that usually complicate the analysis of the QKR~\cite{Shepelyansky:Bicolor:PD87,Rechester:KRDiffCoeff:PRA81,Rechester:Correl:PRL1980}. Recently, it has been shown that a modified and experimentally feasible version of the QKR reproduces the features of this idealized Kicked Rotor~\cite{Hainaut2019IdealNJP}.

\begin{figure}[t!]
	\begin{centering}
		\includegraphics[width=0.45\textwidth]{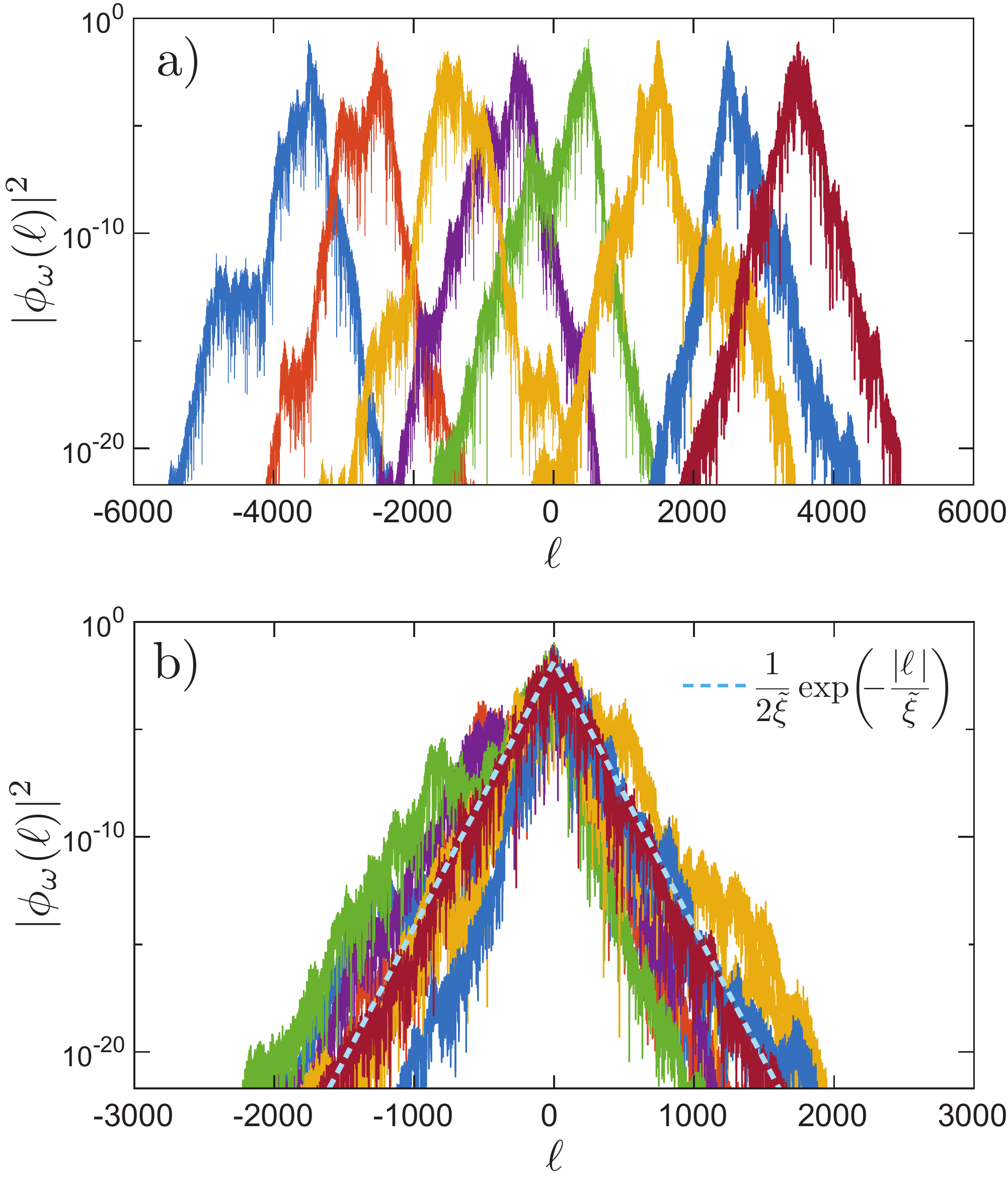}
		\par\end{centering}
	\caption{
		a) Examples of Floquet eigenstates (square modulus, semilog scale), for $K/\kbar=16$.
		b) Same distributions as in a), recentered around $\ell=0$. The dashed line is an exponential distribution with $\tilde{\xi}=35.2$.
	}\label{fig:FigFloquet}
\end{figure}

The Floquet eigenstates $\phi_\omega (\ell)$ of the RKR can be computed by realizing exact diagonalization of the evolution operator $\hat{U}(1)$\footnote{The numerical implementation of the Floquet operator implies periodic boundary conditions in momentum space. Additionally, we choose a cut-off in momentum which is much larger than the localization length of the eigenstates.}.
Fig.~\ref{fig:FigFloquet}.a) shows an example of few such eigenstates, obtained for $K/\kbar=16$.
In order to compare all the Floquet states, we translate them in momentum by a value $\ell_0$ that corresponds to their centroid. Doing so we obtain the distribution presented in Fig.~\ref{fig:FigFloquet}.b) illustrating the fact that they all decay exponentially with similar rates.

\begin{figure}[ht!]
	\begin{centering}
		\includegraphics[width=\linewidth]{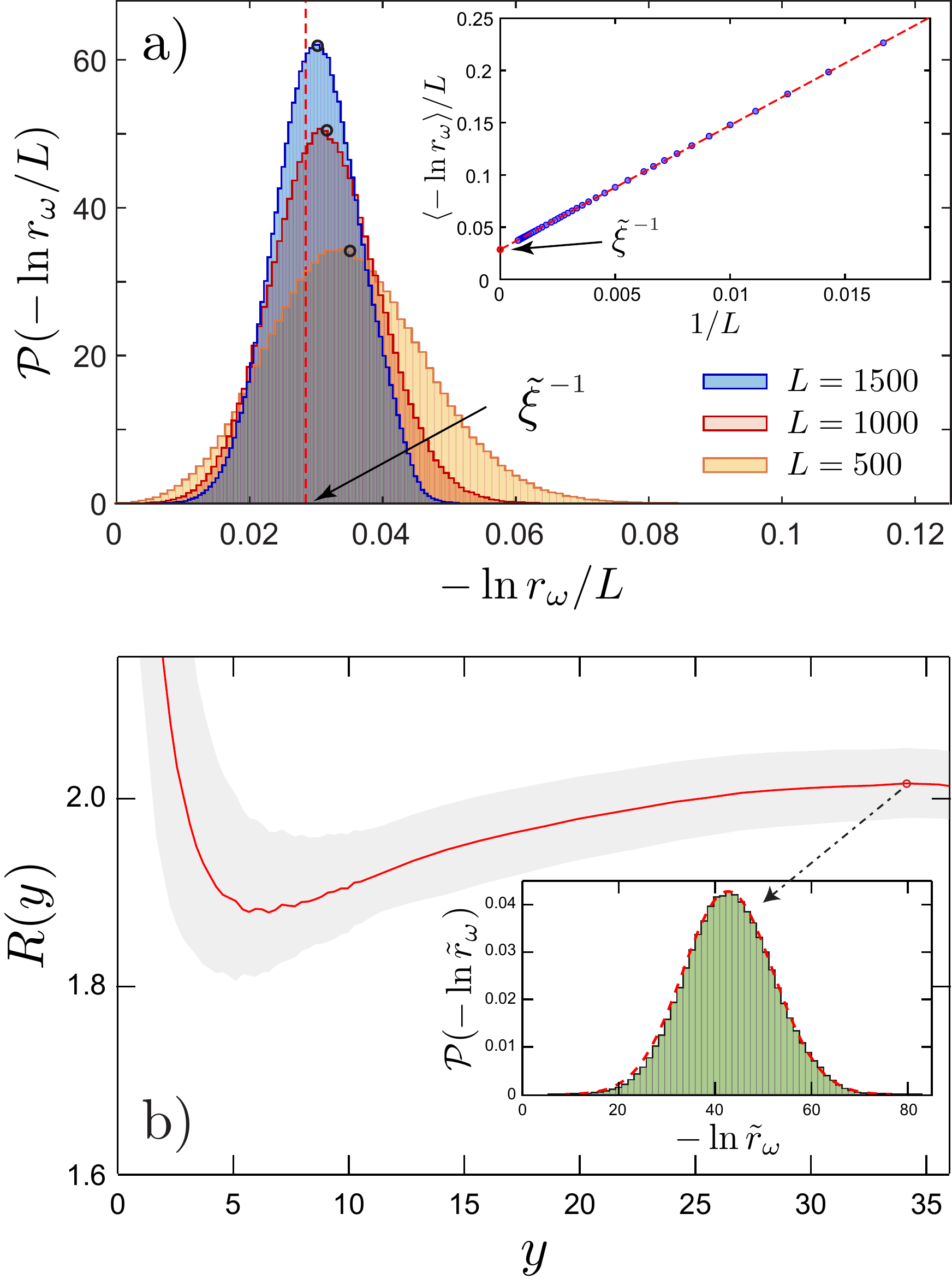}
		\par\end{centering}
	\caption{a) Histograms of the probability distribution of $-\ln(r_\omega)/L$ for various $L$ and $K/\kbar=16$. The circles show the position of the mean, while the vertical dashed line shows the extrapolated mean for $L\to\infty$. The inset shows these means as a function of $1/L$ (symbols), while the line is a fit $f(L)=f_0 + a_0 / L$, used to extrapolate the data and infer $\tilde{\xi}=f_0^{-1}$ ($\tilde{\xi}=35.2\pm 2.1$ in this case). b) Evolution of the ratio $R$ between the variance and mean value of $-\ln \tilde{r}_\omega$, as a function of   $y=L/\tilde{\xi}$, for $K/\kbar=16$. The shaded area represents numerical uncertainty. At large values of $y$, the ratio tends towards a value close to $R = 2$ within the numerical uncertainty, compatible with the prediction of ref.~\cite{Fyodorov1993}.  The inset shows a histogram of the probability distribution of $- \ln \tilde{r}_\omega$, calculated at $y\simeq 35$, which is well fitted by a Gaussian with $R\simeq 2$ (dashed red line). }\label{fig:FigStatMirlin}
\end{figure}

Without loss of generality, we shall study the statistical properties of these `shifted' Floquet states, which possess all the same center implying that $r_\omega(\ell,L)$ now only depends on $L$ : $r_\omega(\ell,L) \mapsto r_{\omega}(L)$. We calculate the histograms of $-\ln r_\omega (L)/L$ for various $L$ using $10^5$ Floquet states with $K/\kbar=16$ and present the results in Fig.~\ref{fig:FigStatMirlin}.a). We see that the distribution gets narrower as $L$ increases. The mean value of each distribution, represented by the circles, slowly converges to $\tilde\xi^{-1}$, as $\propto1/L$, and in practice we can infer the asymptotic value $\tilde{\xi}=(35.2\pm 2.1)$ by extrapolating (via a fit) its $L$-dependence, see the inset of Fig.~\ref{fig:FigStatMirlin}.a). For completeness we plot (dashed line) the distribution corresponding to the obtained average rate $\tilde\xi^{-1}$ in Fig.~\ref{fig:FigFloquet}.b).

To go one step further, we investigate the statistical properties of the dimensionless quantity $\tilde{r}_\omega (y) \equiv 4\tilde{\xi}^2 r_\omega(y)$, with $y\equiv L/\tilde{\xi}$.
First, we compute the ratio between the variance and the mean value of $-\ln \tilde{r}_\omega$:
\begin{equation}
	R(y)=-\frac{\langle \delta^2 (\ln \tilde{r}_{\omega}) \rangle}{\langle \ln \tilde{r}_{\omega} \rangle},
\end{equation}
as a function of $y$. The results are shown in Fig.~\ref{fig:FigStatMirlin}.b). At large $y$ ($L\gg\tilde{\xi}$), we obtain that the ratio $R(y)$ tends to a constant value, close to the theoretical prediction $R\to2$ of Eq.~(\ref{eq:distribPofX}). As shown in the inset, we find that, for $y\gg1$, the probability density of $-\ln \tilde{r}_\omega$ is well fitted by a Gaussian satisfying: $\langle \delta^2 (\ln \tilde{r}_{\omega}) \rangle = -2 \langle \ln \tilde{r}_{\omega} \rangle$, in excellent agreement with Ref.~\cite{Fyodorov1993a}. Similar conclusions have been obtained numerically in~\cite{Pichard1991}  for the conductance fluctuations in quasi-one-dimensional weakly disordered system. 
We have also analyzed the IPR probability distribution, and found a very good agreement with the corresponding supersymmetric predictions, see App.~\ref{app_IPR} for details.

\begin{figure}[t!]
	\begin{centering}
		\includegraphics[width=1.05\linewidth]{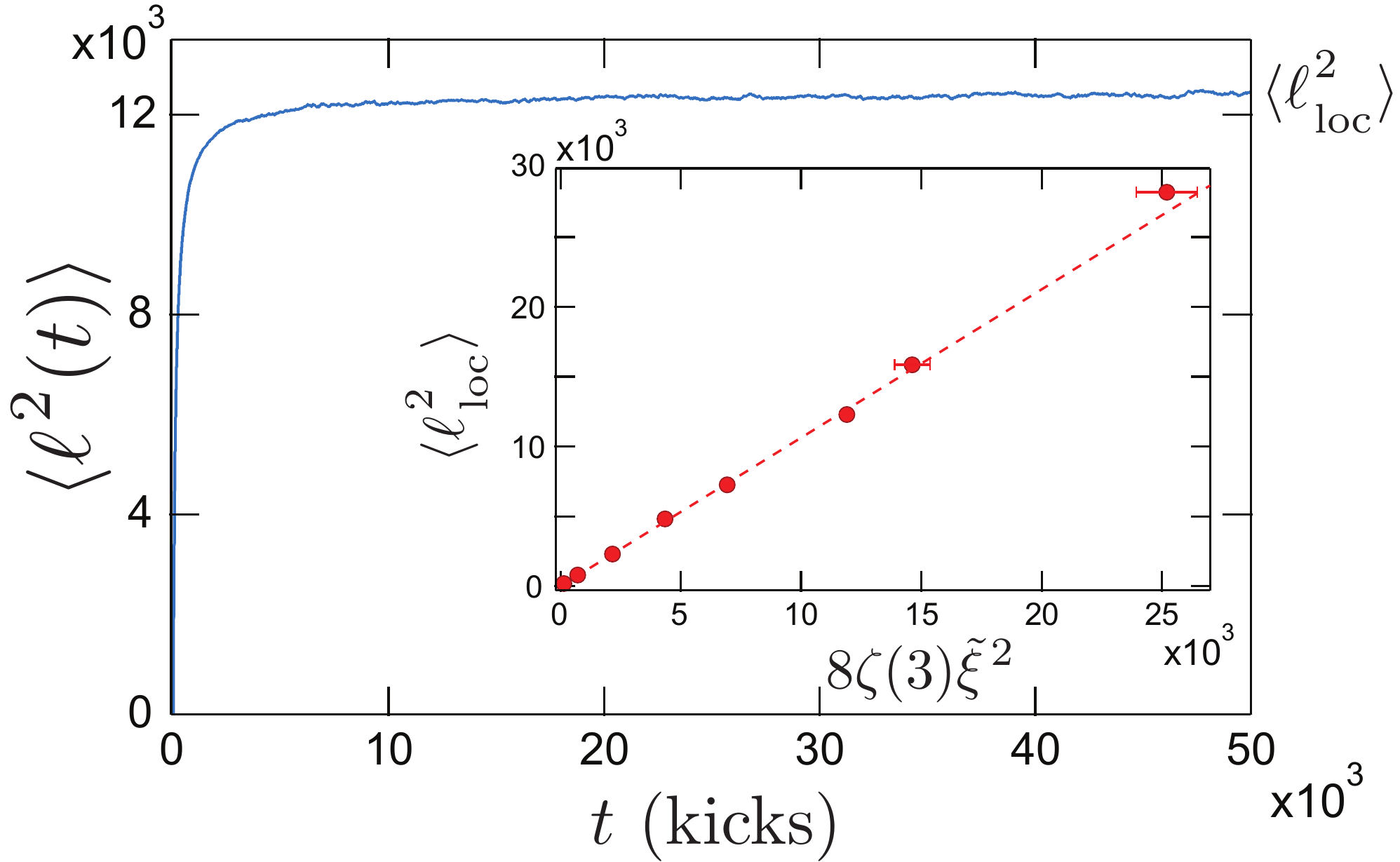}
		\par\end{centering}
	\caption{Time evolution of $\langle \ell^2(t)\rangle$  for $K/\kbar=16$. The steady-state value corresponds to $\langle \ell_{\rm loc}^2\rangle=12.4 \times 10^3$. The inset shows $\langle \ell_{\rm loc}^2\rangle$ vs $8\zeta(3)\tilde{\xi}^2$, which are expected to be equal, for various values of $K/\kbar$.  The horizontal error bars correspond to statistical fitting error, the vertical error bars are smaller than the symbol size. The dashed line is a fit with slope $1.05$. }\label{fig:FigXiLoc}
\end{figure}

These results can be used to analyze the stationary probability distribution of the QKR at long times, deep in the localized regime. This quantity is accessible numerically, as well as in experiments, and is obtained by studying the evolution of a narrow initial momentum distribution after a time much longer than the localization time $t_{\rm loc}$. We first focus on numerical aspects.

Numerical simulations of the dynamics of the (random) Kicked Rotor are straightforward~\cite{Lemarie2009}. The free evolution between two consecutive kicks is diagonal in momentum representation, while the kick operator is diagonal in position representation. Switching between momentum and position representation is easily done through a Fast Fourier Transform. Such a procedure is equivalent to applying the evolution operator $\hat{U}(1)$ once, and one can of course repeat the procedure a sufficient number of times until dynamical localization is reached\footnote{We truncate the momentum basis insuring that the final state has a support much smaller than the momentum cut-off.}. Using this procedure, we compute the evolution of a large number of random phase realizations, and average the resulting momentum distributions.

\begin{figure}[t!]
	\begin{centering}
		\includegraphics[width=\linewidth]{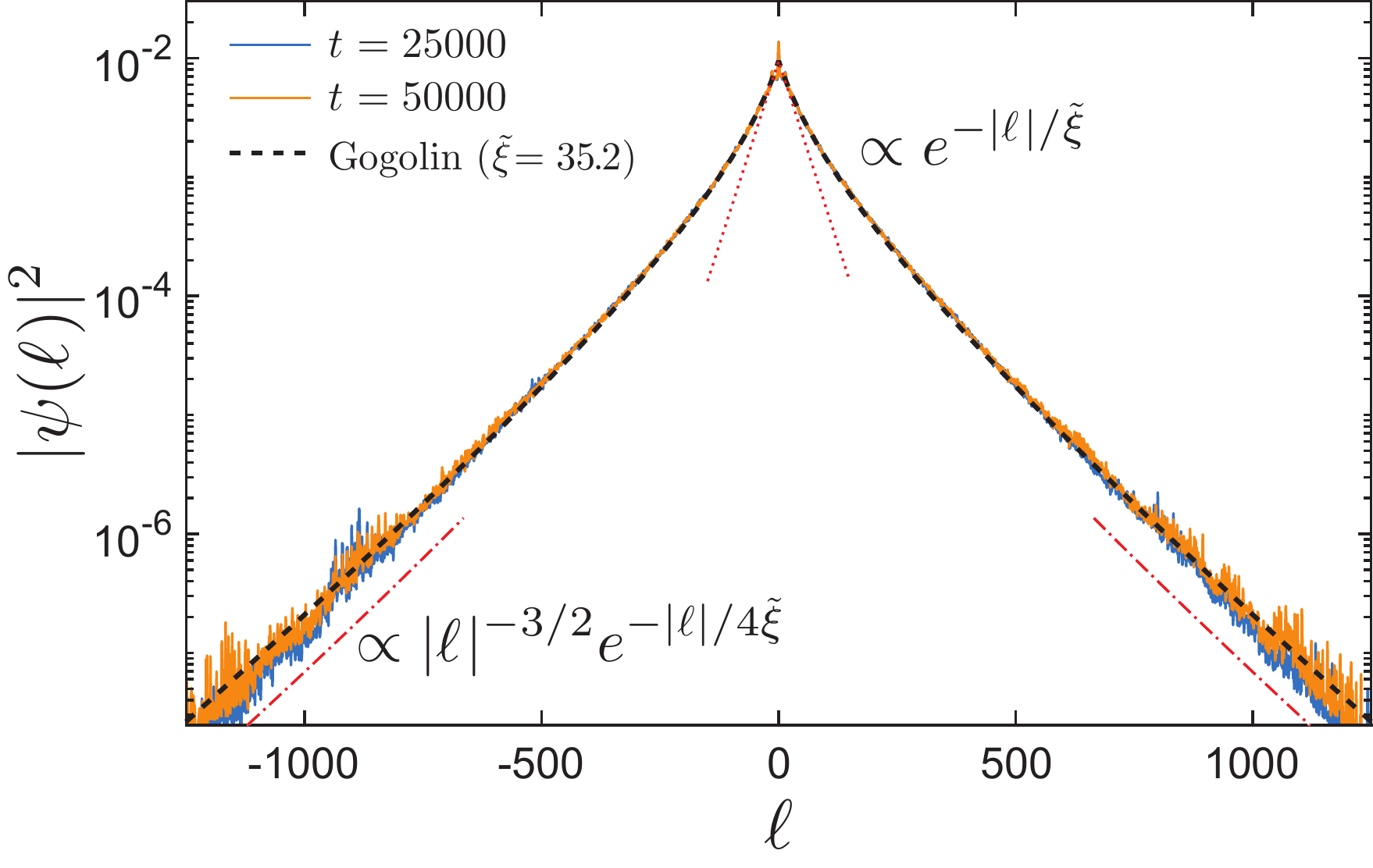}
		\par\end{centering}
	\caption{Steady state momentum distributions at two different times $t\gg t_{\rm loc}$, for $K/\kbar=16$, showing an excellent agreement with the Gogolin distribution (dashed line) with parameter $\xi=35.2=\tilde{\xi}$. The dotted line is the exponential approximation, valid close to $\ell\simeq 0$, whereas the dash-dotted line is the asymptotic limit for $\ell\gg \tilde{\xi}$.}\label{fig:FigGogoRKR}
\end{figure}

We present results for $K/\kbar=16$. At short times the momentum variance $\langle \ell^2(t)\rangle$ grows linearly, then its increase rate slows down on a time scale given by the localization time $t_{\rm loc}$. For $t\gg t_{\rm loc}$, $\langle \ell^2(t)\rangle$ saturates to a value $\langle \ell^2_{\rm loc}\rangle$, see  Fig.~\ref{fig:FigXiLoc}, whereas the momentum distribution becomes stationary. This can be seen in Fig.~\ref{fig:FigGogoRKR}, which shows the momentum distribution at two different times much larger than $t_{\rm loc}$.
We find an excellent agreement with the Gogolin distribution given by Eq.~\eqref{eq:Gogolin}, if we choose $\xi = \tilde{\xi}$, with $\tilde{\xi}$ obtained as described above for the same value of $K/\kbar$, see Fig.~\ref{fig:FigGogoRKR}. This proves that the length scale $\tilde{\xi}$, which characterizes the decay of the Floquet eigenfunctions corresponds \textit{exactly} to the single-parameter $\xi$ which characterizes the functional form of the asymptotic probability density.
Close to the center of the distribution ($\ell \approx 0$), the decay is exponential with a rate $\tilde{\xi}^{-1}$, whereas the large-momentum wings also decrease exponentially (up to an algebraic factor), with a rate four times lower: $1/(4\tilde{\xi})$. This difference is attributed to the strong fluctuations of the $\vert\phi_\omega(l)\vert^2$ \cite{Shepelyansk:KRFloquet:PRL86}, as it is exemplified  in Fig.~\ref{fig:FigFloquet}.

In addition, we checked for different values of the $K/\kbar$ the very good agreement between: 1) the asymptotic momentum distributions  obtained numerically; and 2) the Gogolin distribution with $\xi=\tilde{\xi}$. To be more quantitative, using that $\int \dd\ell \ell^2 \Pi_G(\ell)=8\zeta(3) \xi^2$ \cite{Gogolin1976}, we assess this agreement by comparing $\langle \ell_{\rm loc}^2\rangle$ from our RKR simulations, for a given $K/\kbar$, to  $8\zeta(3)\tilde{\xi}^2$ for the same parameters, see the inset of Fig.~\ref{fig:FigXiLoc}. This shows that the two length scales are equal to within $3\%$.

Finally, we checked the validity of these results for the `standard' QKR (see App.\ref{app_QKR}). It is known that, at low values of $K/\kbar$, the QKR is affected by classical correlation effects, which lead to significant discrepancies with respect to a system with uncorrelated disorder~\cite{Hainaut2019IdealNJP}. At low $K/\kbar$, we found that the classical kick-correlation effects lead to deviations from the predicted Gaussian statistics of Eq.~\ref{eq:distribPofX}. However, the correlation effects disappear at large $K/\kbar$ values, where we find an excellent agreement between the QKR and the ideal RKR model.

\section{Experimental observation of the asymptotic distribution \label{sec_exp}}

The previous section established numerically the relationship between the `intrinsic' localization length $\tilde{\xi}$, characteristic of the exponential decay of the system's eigenfunctions, and the long-time probability distribution, obtained when starting from a peaked initial condition. We shall now focus on the experimental investigation of this characteristic asymptotic shape, and on its distinction from the commonly-thought exponential shape associated with Anderson localization.

In order to access experimentally the question of the exact form of the momentum distribution at long times, it is necessary to realize experimentally an ideal version of the QKR where the late time dynamics is not plagued by correlation effects for experimentally accessible low values of the kick strength. This is rendered possible by using a periodically phase shifted version of the QKR \cite{Tian:EhrenfestTimeDynamicalLoc:PRB05,Hainaut:ERO:PRL17}, described by the Hamiltonian:
\begin{equation}
	\hat H = \frac{\hat p ^2}{2}+K \sum_{n}\cos(\hat x +a_n) \: \delta(t-n).
	\label{Eq:Ham_3Phase}
\end{equation}
For $a_n = 0$, $\forall n$, $\hat H(t)$ reduces to the Hamiltonian of the usual QKR. In this work we will restrict ourselves to a period $N=3$ phase shift ($a_{n+N}=a_n$). We only consider phase shifts such that the Hamiltonian is time-reversal invariant, e.g. $a_1=-a_3$ and $a_2=0$, see \cite{Hainaut2018CFS,Hainaut2019IdealNJP} for details. This insures that our phase-shifted QKR belongs to the same (orthogonal) universality class as the `standard' QKR.

The following experiments are performed by kicking a laser-cooled Cs atomic cloud (temperature $T\simeq 2 \mu$K) using a far-detuned, pulsed optical standing wave (SW), with a period $T_1$. The SW is created by two independent lasers beams, which allows us to control the amplitude and phase of the potential, using the RF signals driving two different acousto-optic modulators. We can thus shape the phase shift sequence $a_n$ at will, and generate the Hamiltonian of Eq.~\eqref{Eq:Ham_3Phase}. The laser parameters are: the detuning $\Delta = - 13$ GHz (at the Cs D2 line, wavelength $\lambda =2\pi/k_L= 852.2$ nm), $1/e$ radius $w_0 = 800$~$\mu$m, and the maximum intensity $I = 30$~W/$\text{cm}^{2}$ for each beam. The pulse duration is $\tau = 200$ ns, while $T_1=9.6$ $\mu$s. From these parameters we get $\kbar=1$ as well as kick amplitudes $K=k_{L}^2 \tau T_1 \hbar I \Gamma^2 / (8M I_{\rm sat}^2 \Delta)$ up to 6 (where $I_{\rm sat}\simeq2.71$~mW/cm$^{2}$ is the saturation intensity and $\Gamma=5.22$~MHz the natural linewidth of the transition). After the desired number of kicks, the cloud expands and the momentum probability density $\Pi(p)=\lvert\Psi(p)\rvert^2 $ is measured using a time-of-flight technique.

\begin{figure}[t!]
	\centering
	\includegraphics[width=1\linewidth]{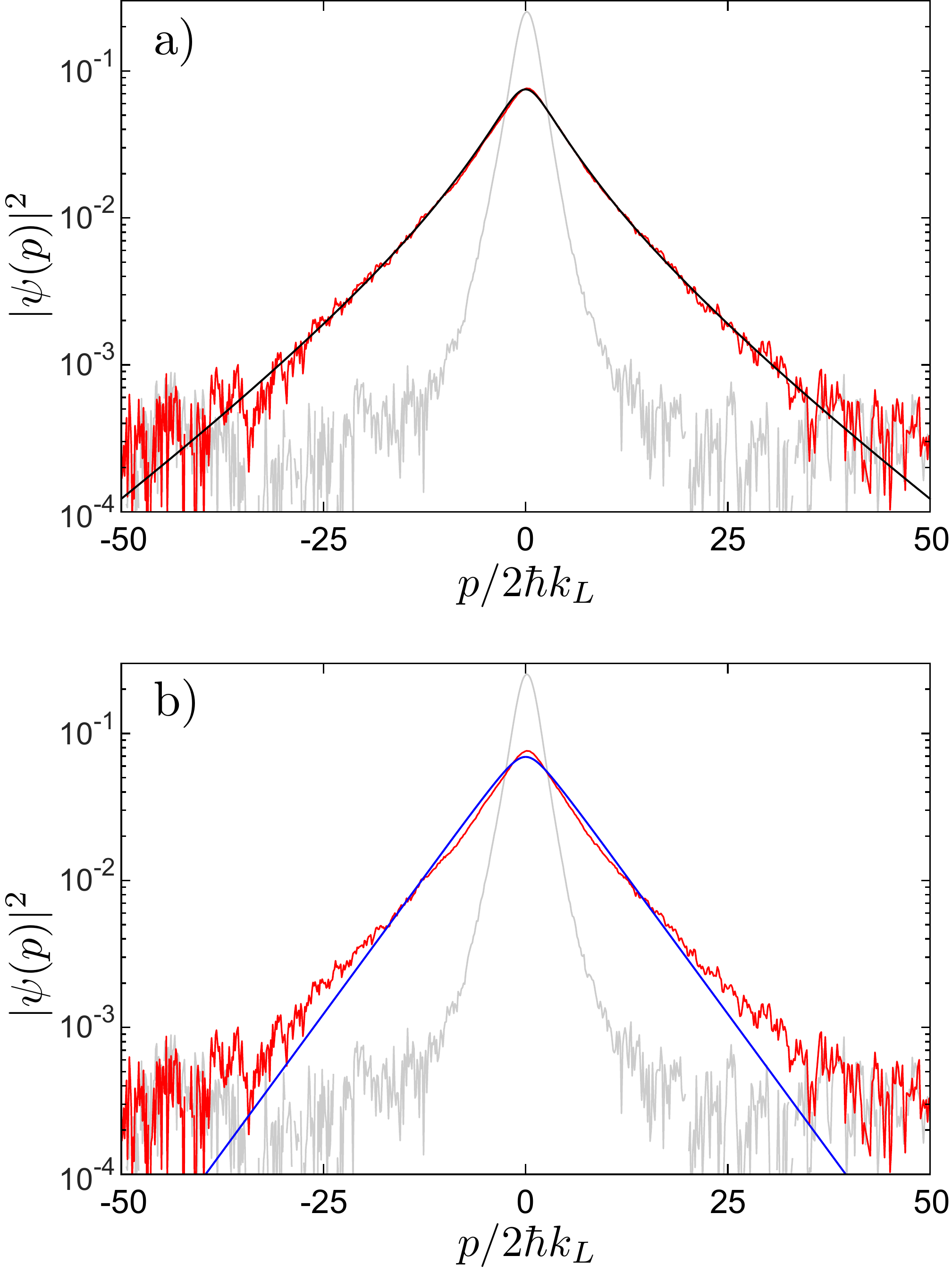}
	\caption{Experimental momentum distribution after $210$ kicks ($K=2.5$, $\kbar=1$) averaged over 100 samplings of triplets $\{a_0,a_1,a_2\}$, and a)  Gogolin fit with $\xi=3.2\times2\hbar k_L$; b) exponential fit with inverse decay rate $5.5\times2\hbar k_L$. Both fits are performed using the theoretical forms convolved with the initial state. The grey curve is the initial momentum distribution of the system.
	}
	\label{Fig:DistribExpLin}
\end{figure}

To realize an accurate analysis of the shape of the final (dynamically-localized) momentum distribution, a careful characterization of the initial state, obtained after the laser cooling stages, is required. Indeed, to perform a meaningful comparison with the theoretical prediction, the initial distribution has to be measured and taken into account through a convolution with the Gogolin distribution. The initial momentum distribution obtained in our experiment is shown in Fig.~\ref{Fig:DistribExpLin}, and is well-approximated with by a Lorentzian shape~\cite{Sortais2000} $D(p)\simeq \frac{2}{\pi\sigma} (1+p^2/\sigma^2)^{-2}$, with $\sigma\simeq 2.31\times2\hbar k_L$ (see App.~\ref{app_exp} for details).

Starting from this initial state, we utilize the Hamiltonian in Eq.~(\ref{Eq:Ham_3Phase}), with parameters $K=2.5$ and $\kbar=1$, and average over 100 realizations of the periodic phase sequences $a_n$. The momentum distribution is measured after a time $t=210$~kicks, and the result is shown in Fig.~\ref{Fig:DistribExpLin}. We have experimentally verified that the distribution reached a steady-state (see App.~\ref{app_exp}, Fig.~\ref{Fig:FigExpeSM} b)), which proves that dynamical localization has been attained. In panel a), we fit the experimental distribution with a Gogolin distribution convolved with the initial distribution, and find a very good agreement both near the center and in the wings, with $\xi=3.2\times2\hbar k_L$.\footnote{The width of the final distribution, given by that of the Gogolin distribution $\sqrt{8\zeta(3)} \xi$, is about four times larger than that of the initial distribution $\sigma$. Therefore, the final distribution is dominated by the localization effects and not the initial state. }
In contrast, panel b) shows a fit of the same final experimental distribution, using an exponential function convolved with the initial state, with an inverse decay rate of $5.5\times 2\hbar k_L$. The exponential shape does not describe well neither the center nor the wings of the experimental data. The ratio of the $\chi^2$ values corresponding to the two fits presented in Fig.~\ref{Fig:DistribExpLin} is $\simeq 0.03$, which clearly proves that the experimental long-time momentum distribution is better described by a Gogolin distribution than by an exponential form.

\section{Conclusion \label{sec_concl}}
In this work, we have investigated the asymptotic properties of a wave-packet localized by disorder and their connection to the statistics of the Floquet eigenstates. Our numerical simulations were found to be in excellent agreement with the supersymmetric predictions.
Using the versatility of the atomic quantum kicked rotor as a quantum simulator of disordered systems, we precisely measured the localized distribution and shown the excellent agreement with the Gogolin prediction.

One important prediction of the supersymmetric formalism is that the Gogolin shape of the localized distribution is preserved in the unitary symmetry class, when time reversal symmetry is broken, though with a doubling of the localization length.
In perspective, this universal feature could 
in principle be investigated experimentally by using kick sequences breaking time reversal symmetry~\cite{Hainaut2018CFS}. Finally, the detailed description of the dynamics, from weak to strong localization, could also be studied experimentally, though no analytical predictions have been devised yet. We leave these challenging problems for future works.

\paragraph*{Acknowledgements-}
The authors thank Dominique Delande, Tony Prat, Nicolas Cherorret and Gabriel Lemari\'e for fruitful discussions. 
This work was supported by Agence Nationale de la Recherche (Grants K-BEC No. ANR-13-BS04-0001-01 and QRITiC I-SITE ULNE/ ANR-16-IDEX-0004 ULNE), the Labex CEMPI (Grant No. ANR-11-LABX-0007-01), Programme Investissements d Avenir under the program ANR-11-IDEX-0002-02, reference ANR-10-LABX-0037-NEXT, and the Ministry of Higher Education and Research, Hauts de France Council and European Regional Development Fund (ERDF) through the Contrat de Projets Etat-Region (CPER Photonics for Society, P4S).
\paragraph*{Contributions-}
CH performed the experiment, under the supervision of RC, with the help of PS, JFC and JCG. CH and RC analyzed the experimental data. AR and RC performed the numerical simulations and their analysis. CH, RC and AR wrote a first version of the manuscript and all authors have contributed to the final version.

\appendix

\section{Probability distribution of the Inverse Participation Ratio (IPR) \label{app_IPR}}

\begin{figure}[h!]
	\begin{centering}
		\includegraphics[width=\linewidth]{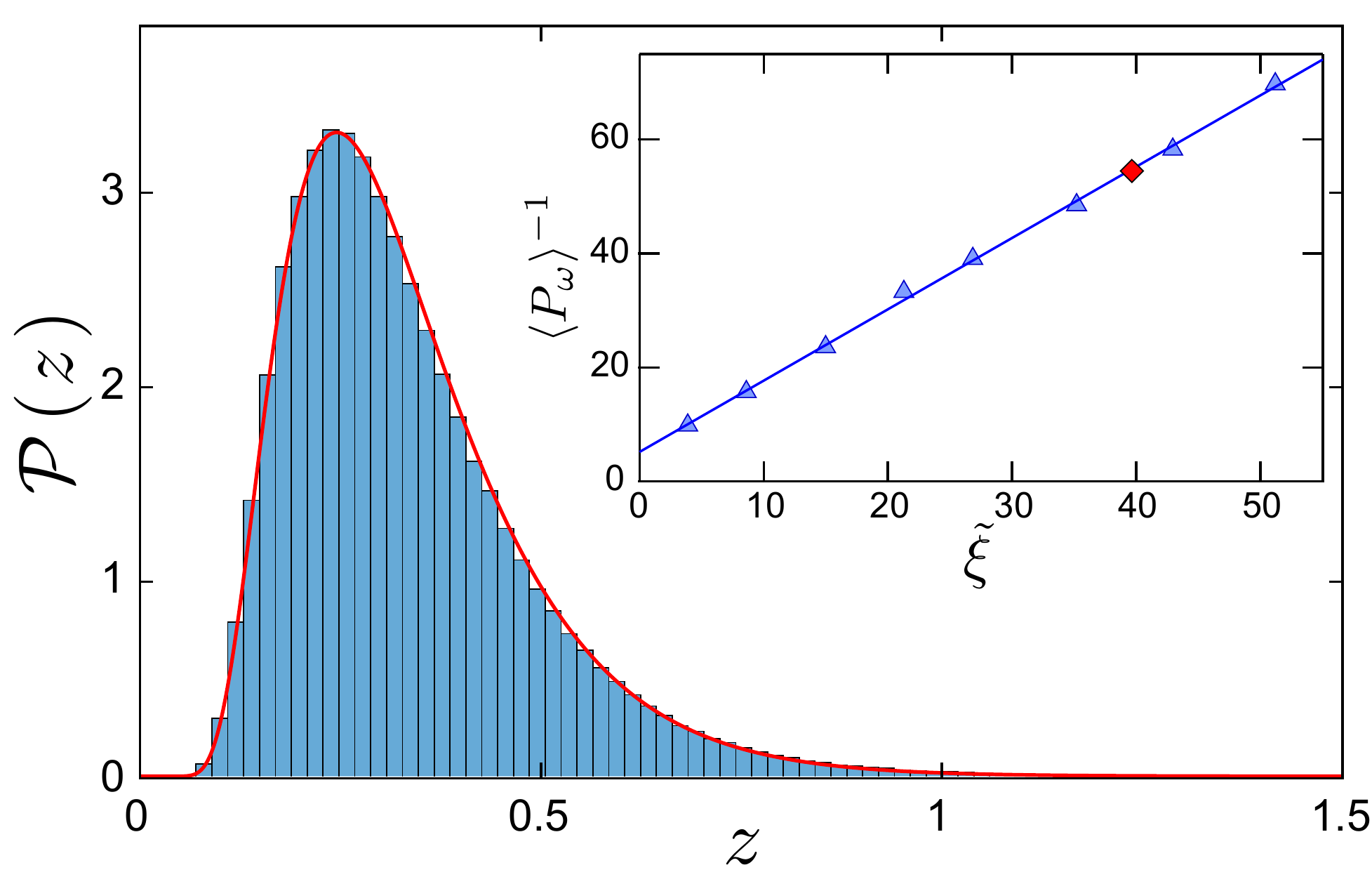}
		\par\end{centering}
	\caption{Histogram of the IPR statistics, computed for the RKR with ($K/\kbar=16$). Red curve: theoretical probability distribution $\mathcal{P}(z)$, given by~Eq.~(\ref{eq:statIPR}). The inset shows the inverse mean IPR $\langle P_\omega \rangle^{-1}$ as a function of $\tilde{\xi}$ for the RKR (triangles). The line is a fit $g(\tilde \xi)=a\tilde\xi+b$, with a slope $a=1.253$ and $b= 5.14$. The diamond corresponds to the same quantity, computed for the QKR ($K=44$ and $\kbar=2.85$), see App.~\ref{app_QKR}.}\label{fig:FigStatIPR}
\end{figure}

To complement the discussion of Sec.~\ref{sec_num}, we analyze here the probability distribution of the IPR. It has been thoroughly studied using supersymmetry, see \cite{Fyodorov1993}.
The IPR of a given Floquet eigenstate $\vert\phi_\omega\rangle$ (at a given $q$) is given by
\begin{equation}
	P_\omega=\sum_\ell \vert\phi_\omega(\ell)\vert^4.
\end{equation}
Defining $z=P_\omega/3\langle P_\omega\rangle$, its probability distribution is given by 
\begin{equation}\label{eq:statIPR}
	\mathcal P(z)=2\pi^2\sum_{k=1}^\infty (2\pi^2 z k^4-3 k^2 )e^{-\pi z k^2}.
\end{equation}
The numerical analysis of the distribution of the IPR for the RKR, similar that described in the main text, gives a very good agreement with the theoretical prediction, as shown in Fig.~\ref{fig:FigStatIPR}. Furthermore, the mean IPR $\langle P_\omega\rangle$ is expected to be a function of $\tilde\xi$ only. This is verified numerically as shown in the inset.

\section{Statistics of eigenstates of the QKR \label{app_QKR}}

In this section, we show that the numerical results obtained in the main text for the RKR are also valid for the standard QKR for large $K/\kbar$, i.e. in a regime where the correlations of the disorder are weak enough.

\begin{figure}[h!]
	\begin{centering}
		\includegraphics[width=\linewidth]{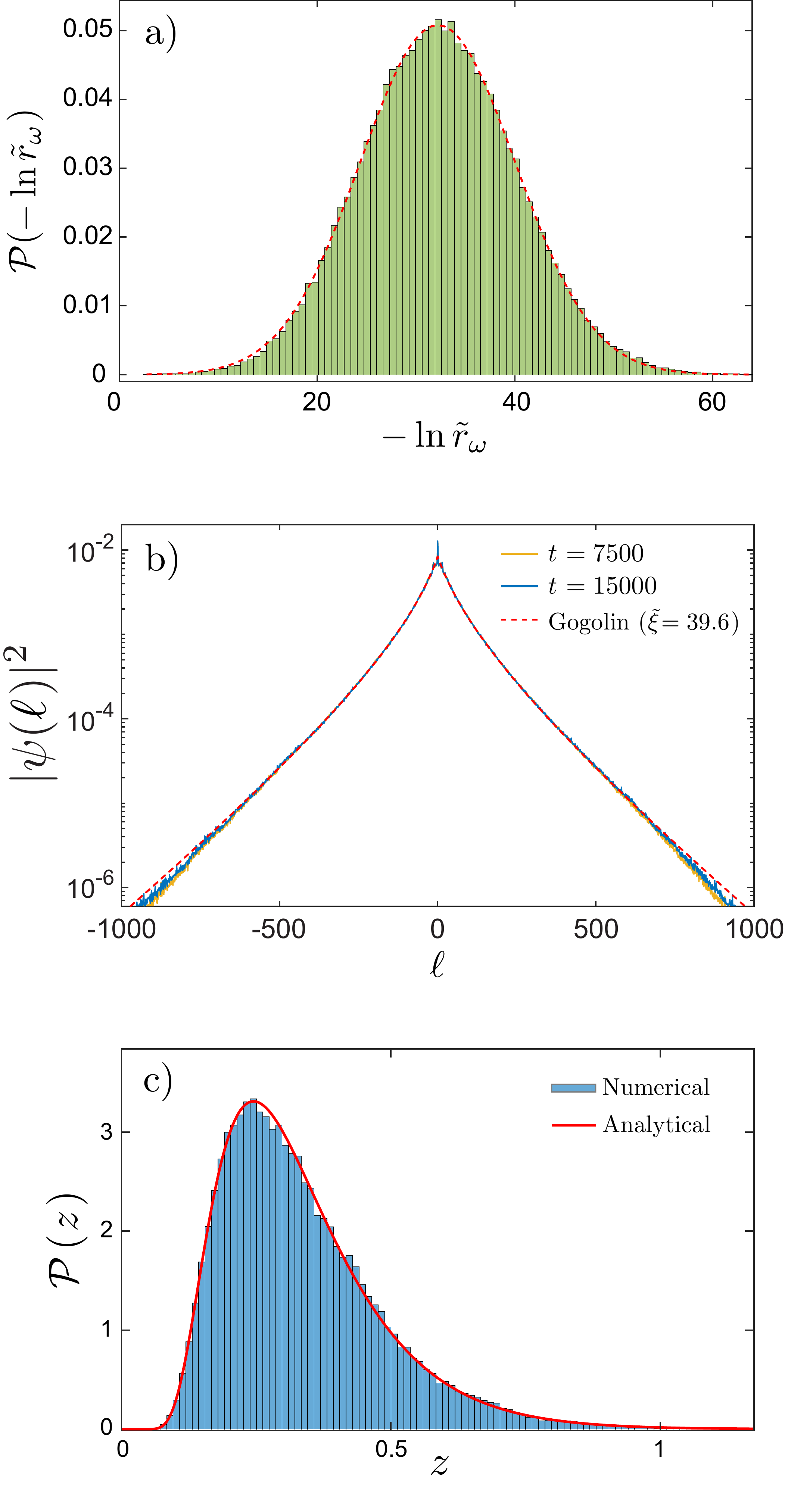}
		\par\end{centering}
	\caption{a) Histogram of the probability distribution of $- \ln \tilde{r}_\omega$, computed for the QKR with $K=44$ and $\kbar=2.85$, and $L/\tilde\xi\simeq 35$. The distribution is well fitted by a Gaussian function (dashed red line). The extrapolation given by Eq.~\eqref{eq:xiOmega} yields $\tilde\xi=39.6\pm3$. b) Corresponding momentum distributions of the QKR, at long times, compared to the Gogolin distribution with $\xi=39.6$. c) Histogram of the IPR statistics, computed for $K=44$ and $\kbar=2.85$, compared to the analytical prediction of Eq.~\eqref{eq:statIPR} (red line).}\label{fig:FigQKRSM}
\end{figure}

Fig.~\ref{fig:FigQKRSM}.a) shows the statistical properties of the Floquet eigenstates computed for $K=44$ and $\kbar=2.85$. We observe that the distribution of $-\ln \tilde{r}_\omega$ is well described by a Gaussian. The corresponding ratio between the variance and the mean, obtained with a Gaussian fit (dashed red line), is $R\simeq1.94 \pm 0.1$, close to the expected value of 2 predicted by Eq.~\eqref{eq:distribPofX}. Following the same procedure as in the RKR case (see Eq.~\eqref{eq:xiOmega}), we obtain an average localization length $\tilde\xi=39.6 \pm 3$. Furthermore, the long-time momentum distribution is given by the Gogolin distribution with $\xi=\tilde\xi$, see Fig~\ref{fig:FigQKRSM}.b).

Finally, the IPR probability distribution of the QKR, shown in Fig~\ref{fig:FigQKRSM}.c), is in very good agreement with Eq.~\eqref{eq:statIPR}. The relation between the IPR mean value and $\tilde\xi$ also agrees with the RKR results (see App.~\ref{app_IPR}, and the diamond data point in the inset of Fig.~\ref{fig:FigStatIPR}).

\section{Experimental initial and final momentum distributions \label{app_exp}}

\begin{figure}[h!]
	\centering
	\includegraphics[width=1\linewidth]{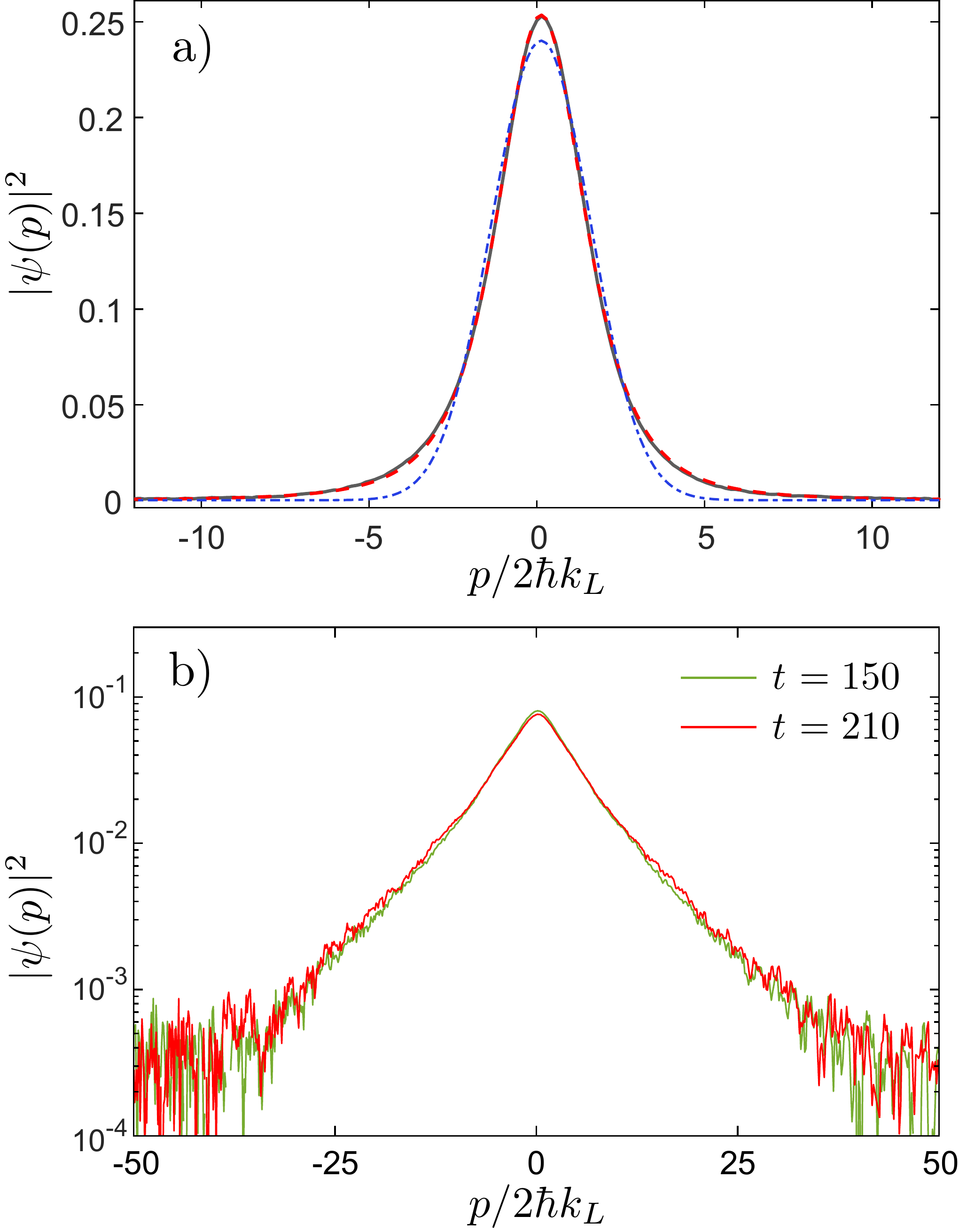}
	\caption{a) Comparison between the initial experimental momentum distribution (full line) with two different fit forms. The fit parameters obtained for the form $D(p)=p_0(1+(p-p_1)^2/p_2^2)^{-p_3}$ are $p_0=0.256/2\hbar k_L$, $p_1=0.098\times 2\hbar k_L$, $p_2=2.311\times 2\hbar k_L$ and $p_3=1.946$ (dashed red line). The Gaussian fit gives a standard deviation of $1.433\times2\hbar k_L$ (dotted-dashed blue line).
		b) Experimental momentum distributions, measured at $t=150$ (green) and $t=210$~kicks (red), showing that the asymptotic stationary state has been reached at these time scales.}
	\label{Fig:FigExpeSM}
\end{figure}

In the experiment, we produce a relatively cold Cs cloud (optical molasses, $T\simeq2$~$\mu$K). As shown in Ref.~\cite{Sortais2000}, this implies that the shape of the momentum distribution differs slightly from a usual Gaussian form. Indeed, when operated near the lower end of the temperature range, the optical molasses momentum distribution displays more weight into its tails, and can be accurately described by a Lorentzian distribution $D(p)=p_0 (1+(p-p_1)^2/p_2^2)^{-p_3}$~\cite{Sortais2000}. We use this functional form, with $p_0$, $p_1$, $p_2$ and $p_3$ as free parameters, to fit the measured momentum distribution of the molasses. As shown in Fig.~\ref{Fig:FigExpeSM}.a), the data are well described by the red dashed-curve corresponding to such distribution. On the other hand, a Gaussian fit (blue dashed curve) clearly shows a significant discrepancy, as it especially underestimates the wings of the initial state distribution. 

Figure~\ref{Fig:FigExpeSM}.b) presents two experimental distributions measured at long times for the parameters mentioned in Sec.~\ref{sec_exp}. The two distributions have barely evolved, showing that the system is indeed deep in the localized regime.

\bibliography{Gogolin-bibliography,bibli_merged}

\begin{thebibliography}{46}%
\makeatletter
\providecommand \@ifxundefined [1]{%
 \@ifx{#1\undefined}
}%
\providecommand \@ifnum [1]{%
 \ifnum #1\expandafter \@firstoftwo
 \else \expandafter \@secondoftwo
 \fi
}%
\providecommand \@ifx [1]{%
 \ifx #1\expandafter \@firstoftwo
 \else \expandafter \@secondoftwo
 \fi
}%
\providecommand \natexlab [1]{#1}%
\providecommand \enquote  [1]{``#1''}%
\providecommand \bibnamefont  [1]{#1}%
\providecommand \bibfnamefont [1]{#1}%
\providecommand \citenamefont [1]{#1}%
\providecommand \href@noop [0]{\@secondoftwo}%
\providecommand \href [0]{\begingroup \@sanitize@url \@href}%
\providecommand \@href[1]{\@@startlink{#1}\@@href}%
\providecommand \@@href[1]{\endgroup#1\@@endlink}%
\providecommand \@sanitize@url [0]{\catcode `\\12\catcode `\$12\catcode
  `\&12\catcode `\#12\catcode `\^12\catcode `\_12\catcode `\%12\relax}%
\providecommand \@@startlink[1]{}%
\providecommand \@@endlink[0]{}%
\providecommand \url  [0]{\begingroup\@sanitize@url \@url }%
\providecommand \@url [1]{\endgroup\@href {#1}{\urlprefix }}%
\providecommand \urlprefix  [0]{URL }%
\providecommand \Eprint [0]{\href }%
\providecommand \doibase [0]{https://doi.org/}%
\providecommand \selectlanguage [0]{\@gobble}%
\providecommand \bibinfo  [0]{\@secondoftwo}%
\providecommand \bibfield  [0]{\@secondoftwo}%
\providecommand \translation [1]{[#1]}%
\providecommand \BibitemOpen [0]{}%
\providecommand \bibitemStop [0]{}%
\providecommand \bibitemNoStop [0]{.\EOS\space}%
\providecommand \EOS [0]{\spacefactor3000\relax}%
\providecommand \BibitemShut  [1]{\csname bibitem#1\endcsname}%
\let\auto@bib@innerbib\@empty
\bibitem [{\citenamefont {Anderson}(1958)}]{Anderson:LocAnderson:PR58}%
  \BibitemOpen
  \bibfield  {author} {\bibinfo {author} {\bibfnamefont {P.~W.}\ \bibnamefont
  {Anderson}},\ }\bibfield  {title} {\bibinfo {title} {{Absence of Diffusion in
  Certain Random Lattices}},\ }\href {https://doi.org/10.1103/PhysRev.109.1492}
  {\bibfield  {journal} {\bibinfo  {journal} {Phys. Rev.}\ }\textbf {\bibinfo
  {volume} {109}},\ \bibinfo {pages} {1492} (\bibinfo {year}
  {1958})}\BibitemShut {NoStop}%
\bibitem [{\citenamefont {Abrahams}\ \emph {et~al.}(1979)\citenamefont
  {Abrahams}, \citenamefont {Anderson}, \citenamefont {Licciardello},\ and\
  \citenamefont {Ramakrishnan}}]{Abrahams:Scaling:PRL79}%
  \BibitemOpen
  \bibfield  {author} {\bibinfo {author} {\bibfnamefont {E.}~\bibnamefont
  {Abrahams}}, \bibinfo {author} {\bibfnamefont {P.~W.}\ \bibnamefont
  {Anderson}}, \bibinfo {author} {\bibfnamefont {D.~C.}\ \bibnamefont
  {Licciardello}},\ and\ \bibinfo {author} {\bibfnamefont {T.~V.}\ \bibnamefont
  {Ramakrishnan}},\ }\bibfield  {title} {\bibinfo {title} {{Scaling Theory of
  Localization\string: Absence of Quantum Diffusion in Two Dimensions}},\
  }\href {https://doi.org/10.1103/PhysRevLett.42.673} {\bibfield  {journal}
  {\bibinfo  {journal} {Phys. Rev. Lett.}\ }\textbf {\bibinfo {volume} {42}},\
  \bibinfo {pages} {673} (\bibinfo {year} {1979})}\BibitemShut {NoStop}%
\bibitem [{\citenamefont {Wiersma}\ \emph {et~al.}(1997)\citenamefont
  {Wiersma}, \citenamefont {Bartolini}, \citenamefont {Lagendijk},\ and\
  \citenamefont {Righini}}]{Wiersma:LightLoc:N97}%
  \BibitemOpen
  \bibfield  {author} {\bibinfo {author} {\bibfnamefont {D.~S.}\ \bibnamefont
  {Wiersma}}, \bibinfo {author} {\bibfnamefont {P.}~\bibnamefont {Bartolini}},
  \bibinfo {author} {\bibfnamefont {A.}~\bibnamefont {Lagendijk}},\ and\
  \bibinfo {author} {\bibfnamefont {R.}~\bibnamefont {Righini}},\ }\bibfield
  {title} {\bibinfo {title} {{Localization of light in a disordered medium}},\
  }\href {https://doi.org/doi:10.1038/37757} {\bibfield  {journal} {\bibinfo
  {journal} {Nature (London)}\ }\textbf {\bibinfo {volume} {390}},\ \bibinfo
  {pages} {671} (\bibinfo {year} {1997})}\BibitemShut {NoStop}%
\bibitem [{\citenamefont {Schwartz}\ \emph {et~al.}(2015)\citenamefont
  {Schwartz}, \citenamefont {Bartal}, \citenamefont {Fishman},\ and\
  \citenamefont {Segev}}]{Schwartz:LocAnderson2DLight:N07}%
  \BibitemOpen
  \bibfield  {author} {\bibinfo {author} {\bibfnamefont {T.}~\bibnamefont
  {Schwartz}}, \bibinfo {author} {\bibfnamefont {G.}~\bibnamefont {Bartal}},
  \bibinfo {author} {\bibfnamefont {S.}~\bibnamefont {Fishman}},\ and\ \bibinfo
  {author} {\bibfnamefont {M.}~\bibnamefont {Segev}},\ }\bibfield  {title}
  {\bibinfo {title} {{Transport and Anderson localization in disordered
  two-dimensional photonic lattices}},\ }\href
  {https://doi.org/10.1038/nature05623} {\bibfield  {journal} {\bibinfo
  {journal} {Nature (London)}\ }\textbf {\bibinfo {volume} {446}},\ \bibinfo
  {pages} {52} (\bibinfo {year} {2015})}\BibitemShut {NoStop}%
\bibitem [{\citenamefont {Dalichaouch}\ \emph {et~al.}(1991)\citenamefont
  {Dalichaouch}, \citenamefont {Armstrong}, \citenamefont {Schultz},
  \citenamefont {Platzman},\ and\ \citenamefont {McCall}}]{Dalichaouch1991}%
  \BibitemOpen
  \bibfield  {author} {\bibinfo {author} {\bibfnamefont {R.}~\bibnamefont
  {Dalichaouch}}, \bibinfo {author} {\bibfnamefont {J.~P.}\ \bibnamefont
  {Armstrong}}, \bibinfo {author} {\bibfnamefont {S.}~\bibnamefont {Schultz}},
  \bibinfo {author} {\bibfnamefont {P.~M.}\ \bibnamefont {Platzman}},\ and\
  \bibinfo {author} {\bibfnamefont {S.~L.}\ \bibnamefont {McCall}},\ }\bibfield
   {title} {\bibinfo {title} {Microwave localization by two-dimensional random
  scattering},\ }\href {https://doi.org/10.1038/354053a0} {\bibfield  {journal}
  {\bibinfo  {journal} {Nature}\ }\textbf {\bibinfo {volume} {354}},\ \bibinfo
  {pages} {53} (\bibinfo {year} {1991})}\BibitemShut {NoStop}%
\bibitem [{\citenamefont {Chabanov}\ \emph {et~al.}(2000)\citenamefont
  {Chabanov}, \citenamefont {Stoytchev},\ and\ \citenamefont
  {Genack}}]{Chabanov2000}%
  \BibitemOpen
  \bibfield  {author} {\bibinfo {author} {\bibfnamefont {A.~A.}\ \bibnamefont
  {Chabanov}}, \bibinfo {author} {\bibfnamefont {M.}~\bibnamefont
  {Stoytchev}},\ and\ \bibinfo {author} {\bibfnamefont {A.~Z.}\ \bibnamefont
  {Genack}},\ }\bibfield  {title} {\bibinfo {title} {Statistical signatures of
  photon localization},\ }\href {https://doi.org/10.1038/35009055} {\bibfield
  {journal} {\bibinfo  {journal} {Nature}\ }\textbf {\bibinfo {volume} {404}},\
  \bibinfo {pages} {850} (\bibinfo {year} {2000})}\BibitemShut {NoStop}%
\bibitem [{\citenamefont {Weaver}(1990)}]{WEAVER1990}%
  \BibitemOpen
  \bibfield  {author} {\bibinfo {author} {\bibfnamefont {R.}~\bibnamefont
  {Weaver}},\ }\bibfield  {title} {\bibinfo {title} {Anderson localization of
  ultrasound},\ }\href
  {https://doi.org/https://doi.org/10.1016/0165-2125(90)90034-2} {\bibfield
  {journal} {\bibinfo  {journal} {Wave Motion}\ }\textbf {\bibinfo {volume}
  {12}},\ \bibinfo {pages} {129} (\bibinfo {year} {1990})}\BibitemShut
  {NoStop}%
\bibitem [{\citenamefont {Akkermans}\ and\ \citenamefont
  {Montambaux}(2011)}]{AkkermansMontambaux:MesoscopicPhysics:11}%
  \BibitemOpen
  \bibfield  {author} {\bibinfo {author} {\bibfnamefont {E.}~\bibnamefont
  {Akkermans}}\ and\ \bibinfo {author} {\bibfnamefont {G.}~\bibnamefont
  {Montambaux}},\ }\href@noop {} {\emph {\bibinfo {title} {{Mesoscopic Physics
  of Electrons and Photons}}}}\ (\bibinfo  {publisher} {{Cambridge University
  Press}},\ \bibinfo {address} {{Cambridge, UK}},\ \bibinfo {year}
  {2011})\BibitemShut {NoStop}%
\bibitem [{\citenamefont {Billy}\ \emph {et~al.}(2008)\citenamefont {Billy},
  \citenamefont {Josse}, \citenamefont {Zuo}, \citenamefont {Bernard},
  \citenamefont {Hambrecht}, \citenamefont {Lugan}, \citenamefont
  {Cl{\'e}ment}, \citenamefont {Sanchez-Palencia}, \citenamefont {Bouyer},\
  and\ \citenamefont {Aspect}}]{Billy:AndersonBEC1D:N08}%
  \BibitemOpen
  \bibfield  {author} {\bibinfo {author} {\bibfnamefont {J.}~\bibnamefont
  {Billy}}, \bibinfo {author} {\bibfnamefont {V.}~\bibnamefont {Josse}},
  \bibinfo {author} {\bibfnamefont {Z.}~\bibnamefont {Zuo}}, \bibinfo {author}
  {\bibfnamefont {A.}~\bibnamefont {Bernard}}, \bibinfo {author} {\bibfnamefont
  {B.}~\bibnamefont {Hambrecht}}, \bibinfo {author} {\bibfnamefont
  {P.}~\bibnamefont {Lugan}}, \bibinfo {author} {\bibfnamefont
  {D.}~\bibnamefont {Cl{\'e}ment}}, \bibinfo {author} {\bibfnamefont
  {L.}~\bibnamefont {Sanchez-Palencia}}, \bibinfo {author} {\bibfnamefont
  {P.}~\bibnamefont {Bouyer}},\ and\ \bibinfo {author} {\bibfnamefont
  {A.}~\bibnamefont {Aspect}},\ }\bibfield  {title} {\bibinfo {title} {{Direct
  observation of Anderson localization of matter-waves in a controlled
  disorder}},\ }\href {https://doi.org/10.1038/nature07000} {\bibfield
  {journal} {\bibinfo  {journal} {Nature (London)}\ }\textbf {\bibinfo {volume}
  {453}},\ \bibinfo {pages} {891} (\bibinfo {year} {2008})}\BibitemShut
  {NoStop}%
\bibitem [{\citenamefont {Chab{\'e}}\ \emph {et~al.}(2008)\citenamefont
  {Chab{\'e}}, \citenamefont {Lemari{\'e}}, \citenamefont {Gr{\'e}maud},
  \citenamefont {Delande}, \citenamefont {Szriftgiser},\ and\ \citenamefont
  {Garreau}}]{Chabe:Anderson:PRL08}%
  \BibitemOpen
  \bibfield  {author} {\bibinfo {author} {\bibfnamefont {J.}~\bibnamefont
  {Chab{\'e}}}, \bibinfo {author} {\bibfnamefont {G.}~\bibnamefont
  {Lemari{\'e}}}, \bibinfo {author} {\bibfnamefont {B.}~\bibnamefont
  {Gr{\'e}maud}}, \bibinfo {author} {\bibfnamefont {D.}~\bibnamefont
  {Delande}}, \bibinfo {author} {\bibfnamefont {P.}~\bibnamefont
  {Szriftgiser}},\ and\ \bibinfo {author} {\bibfnamefont {J.~C.}\ \bibnamefont
  {Garreau}},\ }\bibfield  {title} {\bibinfo {title} {{Experimental Observation
  of the Anderson Metal-Insulator Transition with Atomic Matter Waves}},\
  }\href {https://doi.org/10.1103/PhysRevLett.101.255702} {\bibfield  {journal}
  {\bibinfo  {journal} {Phys. Rev. Lett.}\ }\textbf {\bibinfo {volume} {101}},\
  \bibinfo {pages} {255702} (\bibinfo {year} {2008})}\BibitemShut {NoStop}%
\bibitem [{\citenamefont {{Gogolin}}(1976)}]{Gogolin1976}%
  \BibitemOpen
  \bibfield  {author} {\bibinfo {author} {\bibfnamefont {A.~A.}\ \bibnamefont
  {{Gogolin}}},\ }\bibfield  {title} {\bibinfo {title} {{Electron density
  distribution for localized states in a one-dimensional disordered system}},\
  }\href@noop {} {\bibfield  {journal} {\bibinfo  {journal} {Soviet Journal of
  Experimental and Theoretical Physics}\ }\textbf {\bibinfo {volume} {44}},\
  \bibinfo {pages} {1003} (\bibinfo {year} {1976})}\BibitemShut {NoStop}%
\bibitem [{\citenamefont {{Gogolin}}\ \emph {et~al.}(1975)\citenamefont
  {{Gogolin}}, \citenamefont {{Mel'Nikov}},\ and\ \citenamefont
  {{Rashba}}}]{Gogolin1975}%
  \BibitemOpen
  \bibfield  {author} {\bibinfo {author} {\bibfnamefont {A.~A.}\ \bibnamefont
  {{Gogolin}}}, \bibinfo {author} {\bibfnamefont {V.~I.}\ \bibnamefont
  {{Mel'Nikov}}},\ and\ \bibinfo {author} {\bibfnamefont {{\'E}.~I.}\
  \bibnamefont {{Rashba}}},\ }\bibfield  {title} {\bibinfo {title}
  {{Conductivity in a disordered one-dimensional system induced by
  electron-phonon interaction}},\ }\href@noop {} {\bibfield  {journal}
  {\bibinfo  {journal} {Soviet Journal of Experimental and Theoretical
  Physics}\ }\textbf {\bibinfo {volume} {42}},\ \bibinfo {pages} {168}
  (\bibinfo {year} {1975})}\BibitemShut {NoStop}%
\bibitem [{\citenamefont {Efetov}(1997)}]{Efetov:SupersymmetryInDisorder:97}%
  \BibitemOpen
  \bibfield  {author} {\bibinfo {author} {\bibfnamefont {K.}~\bibnamefont
  {Efetov}},\ }\href@noop {} {\emph {\bibinfo {title} {{Supersymmetry in
  Disorder and Chaos}}}}\ (\bibinfo  {publisher} {{Cambridge University
  Press}},\ \bibinfo {address} {{Cambridge, UK}},\ \bibinfo {year}
  {1997})\BibitemShut {NoStop}%
\bibitem [{\citenamefont {Efetov}\ and\ \citenamefont
  {Larkin}(1983)}]{Efetov1983a}%
  \BibitemOpen
  \bibfield  {author} {\bibinfo {author} {\bibfnamefont {K.}~\bibnamefont
  {Efetov}}\ and\ \bibinfo {author} {\bibfnamefont {A.}~\bibnamefont
  {Larkin}},\ }\bibfield  {title} {\bibinfo {title} {Kinetics of a quantum
  particle in a long metallic wire},\ }\href@noop {} {\bibfield  {journal}
  {\bibinfo  {journal} {JETP}\ }\textbf {\bibinfo {volume} {58}},\ \bibinfo
  {pages} {764} (\bibinfo {year} {1983})}\BibitemShut {NoStop}%
\bibitem [{\citenamefont {Fyodorov}\ and\ \citenamefont
  {Mirlin}(1991)}]{Fyodorov1991}%
  \BibitemOpen
  \bibfield  {author} {\bibinfo {author} {\bibfnamefont {Y.~V.}\ \bibnamefont
  {Fyodorov}}\ and\ \bibinfo {author} {\bibfnamefont {A.~D.}\ \bibnamefont
  {Mirlin}},\ }\bibfield  {title} {\bibinfo {title} {Scaling properties of
  localization in random band matrices: A \ensuremath{\sigma}-model approach},\
  }\href {https://doi.org/10.1103/PhysRevLett.67.2405} {\bibfield  {journal}
  {\bibinfo  {journal} {Phys. Rev. Lett.}\ }\textbf {\bibinfo {volume} {67}},\
  \bibinfo {pages} {2405} (\bibinfo {year} {1991})}\BibitemShut {NoStop}%
\bibitem [{\citenamefont {Altland}\ and\ \citenamefont
  {Zirnbauer}(1996)}]{Altland1996}%
  \BibitemOpen
  \bibfield  {author} {\bibinfo {author} {\bibfnamefont {A.}~\bibnamefont
  {Altland}}\ and\ \bibinfo {author} {\bibfnamefont {M.~R.}\ \bibnamefont
  {Zirnbauer}},\ }\bibfield  {title} {\bibinfo {title} {Field theory of the
  quantum kicked rotor},\ }\href {https://doi.org/10.1103/PhysRevLett.77.4536}
  {\bibfield  {journal} {\bibinfo  {journal} {Phys. Rev. Lett.}\ }\textbf
  {\bibinfo {volume} {77}},\ \bibinfo {pages} {4536} (\bibinfo {year}
  {1996})}\BibitemShut {NoStop}%
\bibitem [{\citenamefont {Izrailev}(1990)}]{Izrailev:LocDyn:PREP90}%
  \BibitemOpen
  \bibfield  {author} {\bibinfo {author} {\bibfnamefont {F.~M.}\ \bibnamefont
  {Izrailev}},\ }\bibfield  {title} {\bibinfo {title} {{Simple models of
  quantum chaos\string: spectrum and eigenfunctions}},\ }\href@noop {}
  {\bibfield  {journal} {\bibinfo  {journal} {Phys. Rep.}\ }\textbf {\bibinfo
  {volume} {196}},\ \bibinfo {pages} {299} (\bibinfo {year}
  {1990})}\BibitemShut {NoStop}%
\bibitem [{\citenamefont {Casati}\ \emph {et~al.}(1979)\citenamefont {Casati},
  \citenamefont {Chirikov}, \citenamefont {Ford},\ and\ \citenamefont
  {Izrailev}}]{Casati:LocDynFirst:LNP79}%
  \BibitemOpen
  \bibfield  {author} {\bibinfo {author} {\bibfnamefont {G.}~\bibnamefont
  {Casati}}, \bibinfo {author} {\bibfnamefont {B.~V.}\ \bibnamefont
  {Chirikov}}, \bibinfo {author} {\bibfnamefont {J.}~\bibnamefont {Ford}},\
  and\ \bibinfo {author} {\bibfnamefont {F.~M.}\ \bibnamefont {Izrailev}},\
  }\bibinfo {title} {{Stochastic behavior of a quantum pendulum under a
  periodic perturbation}},\ in\ \href {https://doi.org/10.1007/BFb0021757}
  {\emph {\bibinfo {booktitle} {{Stochastic Behavior in Classical and Quantum
  Hamiltonian Systems: Volta Memorial Conference, Como, 1977}}}},\
  Vol.~\bibinfo {volume} {93},\ \bibinfo {editor} {edited by\ \bibinfo {editor}
  {\bibnamefont {{G. Casati and J. Ford}}}}\ (\bibinfo  {publisher} {{Springer
  Berlin Heidelberg}},\ \bibinfo {address} {{Berlin, Germany}},\ \bibinfo
  {year} {1979})\ pp.\ \bibinfo {pages} {334--352}\BibitemShut {NoStop}%
\bibitem [{\citenamefont {Fishman}\ \emph {et~al.}(1982)\citenamefont
  {Fishman}, \citenamefont {Grempel},\ and\ \citenamefont
  {Prange}}]{Fishman:LocDynAnders:PRL82}%
  \BibitemOpen
  \bibfield  {author} {\bibinfo {author} {\bibfnamefont {S.}~\bibnamefont
  {Fishman}}, \bibinfo {author} {\bibfnamefont {D.~R.}\ \bibnamefont
  {Grempel}},\ and\ \bibinfo {author} {\bibfnamefont {R.~E.}\ \bibnamefont
  {Prange}},\ }\bibfield  {title} {\bibinfo {title} {{Chaos, Quantum
  Recurrences, and Anderson Localization}},\ }\href
  {https://doi.org/10.1103/PhysRevLett.49.509} {\bibfield  {journal} {\bibinfo
  {journal} {Phys. Rev. Lett.}\ }\textbf {\bibinfo {volume} {49}},\ \bibinfo
  {pages} {509} (\bibinfo {year} {1982})}\BibitemShut {NoStop}%
\bibitem [{\citenamefont {Garreau}(2017)}]{Garreau2017}%
  \BibitemOpen
  \bibfield  {author} {\bibinfo {author} {\bibfnamefont {J.-C.}\ \bibnamefont
  {Garreau}},\ }\bibfield  {title} {\bibinfo {title} {Quantum simulation of
  disordered systems with cold atoms},\ }\href
  {https://doi.org/https://doi.org/10.1016/j.crhy.2016.09.002} {\bibfield
  {journal} {\bibinfo  {journal} {Comptes Rendus Physique}\ }\textbf {\bibinfo
  {volume} {18}},\ \bibinfo {pages} {31} (\bibinfo {year} {2017})},\ \bibinfo
  {note} {prizes of the French Academy of Sciences 2015 / Prix de l'Acad\'emie
  des sciences 2015}\BibitemShut {NoStop}%
\bibitem [{\citenamefont {Moore}\ \emph {et~al.}(1995)\citenamefont {Moore},
  \citenamefont {Robinson}, \citenamefont {Bharucha}, \citenamefont
  {Sundaram},\ and\ \citenamefont
  {Raizen}}]{Moore:AtomOpticsRealizationQKR:PRL95}%
  \BibitemOpen
  \bibfield  {author} {\bibinfo {author} {\bibfnamefont {F.~L.}\ \bibnamefont
  {Moore}}, \bibinfo {author} {\bibfnamefont {J.~C.}\ \bibnamefont {Robinson}},
  \bibinfo {author} {\bibfnamefont {C.~F.}\ \bibnamefont {Bharucha}}, \bibinfo
  {author} {\bibfnamefont {B.}~\bibnamefont {Sundaram}},\ and\ \bibinfo
  {author} {\bibfnamefont {M.~G.}\ \bibnamefont {Raizen}},\ }\bibfield  {title}
  {\bibinfo {title} {{Atom Optics Realization of the Quantum $\delta$-Kicked
  Rotor}},\ }\href {https://doi.org/10.1103/PhysRevLett.75.4598} {\bibfield
  {journal} {\bibinfo  {journal} {Phys. Rev. Lett.}\ }\textbf {\bibinfo
  {volume} {75}},\ \bibinfo {pages} {4598} (\bibinfo {year}
  {1995})}\BibitemShut {NoStop}%
\bibitem [{\citenamefont {Moore}\ \emph {et~al.}(1994)\citenamefont {Moore},
  \citenamefont {Robinson}, \citenamefont {Bharucha}, \citenamefont
  {Williams},\ and\ \citenamefont {Raizen}}]{Moore:LDynFirst:PRL94}%
  \BibitemOpen
  \bibfield  {author} {\bibinfo {author} {\bibfnamefont {F.~L.}\ \bibnamefont
  {Moore}}, \bibinfo {author} {\bibfnamefont {J.~C.}\ \bibnamefont {Robinson}},
  \bibinfo {author} {\bibfnamefont {C.}~\bibnamefont {Bharucha}}, \bibinfo
  {author} {\bibfnamefont {P.~E.}\ \bibnamefont {Williams}},\ and\ \bibinfo
  {author} {\bibfnamefont {M.~G.}\ \bibnamefont {Raizen}},\ }\bibfield  {title}
  {\bibinfo {title} {{Observation of Dynamical Localization in Atomic Momentum
  Transfer\string: A New Testing Ground for Quantum Chaos}},\ }\href
  {https://doi.org/10.1103/PhysRevLett.73.2974} {\bibfield  {journal} {\bibinfo
   {journal} {Phys. Rev. Lett.}\ }\textbf {\bibinfo {volume} {73}},\ \bibinfo
  {pages} {2974} (\bibinfo {year} {1994})}\BibitemShut {NoStop}%
\bibitem [{\citenamefont {Manai}\ \emph {et~al.}(2015)\citenamefont {Manai},
  \citenamefont {Cl{\'e}ment}, \citenamefont {Chicireanu}, \citenamefont
  {Hainaut}, \citenamefont {Garreau}, \citenamefont {Szriftgiser},\ and\
  \citenamefont {Delande}}]{Manai:Anderson2DKR:PRL15}%
  \BibitemOpen
  \bibfield  {author} {\bibinfo {author} {\bibfnamefont {I.}~\bibnamefont
  {Manai}}, \bibinfo {author} {\bibfnamefont {J.-F.}\ \bibnamefont
  {Cl{\'e}ment}}, \bibinfo {author} {\bibfnamefont {R.}~\bibnamefont
  {Chicireanu}}, \bibinfo {author} {\bibfnamefont {C.}~\bibnamefont {Hainaut}},
  \bibinfo {author} {\bibfnamefont {J.~C.}\ \bibnamefont {Garreau}}, \bibinfo
  {author} {\bibfnamefont {P.}~\bibnamefont {Szriftgiser}},\ and\ \bibinfo
  {author} {\bibfnamefont {D.}~\bibnamefont {Delande}},\ }\bibfield  {title}
  {\bibinfo {title} {{Experimental Observation of Two-Dimensional Anderson
  Localization with the Atomic Kicked Rotor}},\ }\href
  {https://doi.org/10.1103/PhysRevLett.115.240603} {\bibfield  {journal}
  {\bibinfo  {journal} {Phys. Rev. Lett.}\ }\textbf {\bibinfo {volume} {115}},\
  \bibinfo {pages} {240603} (\bibinfo {year} {2015})}\BibitemShut {NoStop}%
\bibitem [{\citenamefont {Lopez}\ \emph {et~al.}(2012)\citenamefont {Lopez},
  \citenamefont {Cl{\'e}ment}, \citenamefont {Szriftgiser}, \citenamefont
  {Garreau},\ and\ \citenamefont
  {Delande}}]{Lopez:ExperimentalTestOfUniversality:PRL12}%
  \BibitemOpen
  \bibfield  {author} {\bibinfo {author} {\bibfnamefont {M.}~\bibnamefont
  {Lopez}}, \bibinfo {author} {\bibfnamefont {J.-F.}\ \bibnamefont
  {Cl{\'e}ment}}, \bibinfo {author} {\bibfnamefont {P.}~\bibnamefont
  {Szriftgiser}}, \bibinfo {author} {\bibfnamefont {J.~C.}\ \bibnamefont
  {Garreau}},\ and\ \bibinfo {author} {\bibfnamefont {D.}~\bibnamefont
  {Delande}},\ }\bibfield  {title} {\bibinfo {title} {{Experimental Test of
  Universality of the Anderson Transition}},\ }\href
  {https://doi.org/10.1103/PhysRevLett.108.095701} {\bibfield  {journal}
  {\bibinfo  {journal} {Phys. Rev. Lett.}\ }\textbf {\bibinfo {volume} {108}},\
  \bibinfo {pages} {095701} (\bibinfo {year} {2012})}\BibitemShut {NoStop}%
\bibitem [{\citenamefont {Hainaut}\ \emph {et~al.}(2017)\citenamefont
  {Hainaut}, \citenamefont {Manai}, \citenamefont {Chicireanu}, \citenamefont
  {Cl\'ement}, \citenamefont {Zemmouri}, \citenamefont {Garreau}, \citenamefont
  {Szriftgiser}, \citenamefont {Lemari\'e}, \citenamefont {Cherroret},\ and\
  \citenamefont {Delande}}]{Hainaut:ERO:PRL17}%
  \BibitemOpen
  \bibfield  {author} {\bibinfo {author} {\bibfnamefont {C.}~\bibnamefont
  {Hainaut}}, \bibinfo {author} {\bibfnamefont {I.}~\bibnamefont {Manai}},
  \bibinfo {author} {\bibfnamefont {R.}~\bibnamefont {Chicireanu}}, \bibinfo
  {author} {\bibfnamefont {J.-F. m.~c.}\ \bibnamefont {Cl\'ement}}, \bibinfo
  {author} {\bibfnamefont {S.}~\bibnamefont {Zemmouri}}, \bibinfo {author}
  {\bibfnamefont {J.~C.}\ \bibnamefont {Garreau}}, \bibinfo {author}
  {\bibfnamefont {P.}~\bibnamefont {Szriftgiser}}, \bibinfo {author}
  {\bibfnamefont {G.}~\bibnamefont {Lemari\'e}}, \bibinfo {author}
  {\bibfnamefont {N.}~\bibnamefont {Cherroret}},\ and\ \bibinfo {author}
  {\bibfnamefont {D.}~\bibnamefont {Delande}},\ }\bibfield  {title} {\bibinfo
  {title} {Return to the origin as a probe of atomic phase coherence},\ }\href
  {https://doi.org/10.1103/PhysRevLett.118.184101} {\bibfield  {journal}
  {\bibinfo  {journal} {Phys. Rev. Lett.}\ }\textbf {\bibinfo {volume} {118}},\
  \bibinfo {pages} {184101} (\bibinfo {year} {2017})},\ \Eprint
  {https://arxiv.org/abs/1606.07237} {1606.07237} \BibitemShut {NoStop}%
\bibitem [{\citenamefont {Hainaut}\ \emph
  {et~al.}(2018{\natexlab{a}})\citenamefont {Hainaut}, \citenamefont {Manai},
  \citenamefont {Cl{\'e}ment}, \citenamefont {Garreau}, \citenamefont
  {Szriftgiser}, \citenamefont {Lemari{\'e}}, \citenamefont {Cherroret},
  \citenamefont {Delande},\ and\ \citenamefont {Chicireanu}}]{Hainaut2018CFS}%
  \BibitemOpen
  \bibfield  {author} {\bibinfo {author} {\bibfnamefont {C.}~\bibnamefont
  {Hainaut}}, \bibinfo {author} {\bibfnamefont {I.}~\bibnamefont {Manai}},
  \bibinfo {author} {\bibfnamefont {J.-F.}\ \bibnamefont {Cl{\'e}ment}},
  \bibinfo {author} {\bibfnamefont {J.~C.}\ \bibnamefont {Garreau}}, \bibinfo
  {author} {\bibfnamefont {P.}~\bibnamefont {Szriftgiser}}, \bibinfo {author}
  {\bibfnamefont {G.}~\bibnamefont {Lemari{\'e}}}, \bibinfo {author}
  {\bibfnamefont {N.}~\bibnamefont {Cherroret}}, \bibinfo {author}
  {\bibfnamefont {D.}~\bibnamefont {Delande}},\ and\ \bibinfo {author}
  {\bibfnamefont {R.}~\bibnamefont {Chicireanu}},\ }\bibfield  {title}
  {\bibinfo {title} {Controlling symmetry and localization with an artificial
  gauge field in a disordered quantum system},\ }\href
  {https://doi.org/10.1038/s41467-018-03481-9} {\bibfield  {journal} {\bibinfo
  {journal} {Nature Communications}\ }\textbf {\bibinfo {volume} {9}},\
  \bibinfo {pages} {1382} (\bibinfo {year} {2018}{\natexlab{a}})}\BibitemShut
  {NoStop}%
\bibitem [{\citenamefont {Hainaut}\ \emph
  {et~al.}(2018{\natexlab{b}})\citenamefont {Hainaut}, \citenamefont
  {Ran\c{c}on}, \citenamefont {Cl\'ement}, \citenamefont {Garreau},
  \citenamefont {Szriftgiser}, \citenamefont {Chicireanu},\ and\ \citenamefont
  {Delande}}]{Hainaut2018_Ratchet}%
  \BibitemOpen
  \bibfield  {author} {\bibinfo {author} {\bibfnamefont {C.}~\bibnamefont
  {Hainaut}}, \bibinfo {author} {\bibfnamefont {A.}~\bibnamefont {Ran\c{c}on}},
  \bibinfo {author} {\bibfnamefont {J.-F.}\ \bibnamefont {Cl\'ement}}, \bibinfo
  {author} {\bibfnamefont {J.~C.}\ \bibnamefont {Garreau}}, \bibinfo {author}
  {\bibfnamefont {P.}~\bibnamefont {Szriftgiser}}, \bibinfo {author}
  {\bibfnamefont {R.}~\bibnamefont {Chicireanu}},\ and\ \bibinfo {author}
  {\bibfnamefont {D.}~\bibnamefont {Delande}},\ }\bibfield  {title} {\bibinfo
  {title} {Ratchet effect in the quantum kicked rotor and its destruction by
  dynamical localization},\ }\href {https://doi.org/10.1103/PhysRevA.97.061601}
  {\bibfield  {journal} {\bibinfo  {journal} {Phys. Rev. A}\ }\textbf {\bibinfo
  {volume} {97}},\ \bibinfo {pages} {061601} (\bibinfo {year}
  {2018}{\natexlab{b}})}\BibitemShut {NoStop}%
\bibitem [{\citenamefont {Hainaut}\ \emph
  {et~al.}(2018{\natexlab{c}})\citenamefont {Hainaut}, \citenamefont {Fang},
  \citenamefont {Ran\c{c}on}, \citenamefont {Cl\'ement}, \citenamefont
  {Szriftgiser}, \citenamefont {Garreau}, \citenamefont {Tian},\ and\
  \citenamefont {Chicireanu}}]{Hainaut2018}%
  \BibitemOpen
  \bibfield  {author} {\bibinfo {author} {\bibfnamefont {C.}~\bibnamefont
  {Hainaut}}, \bibinfo {author} {\bibfnamefont {P.}~\bibnamefont {Fang}},
  \bibinfo {author} {\bibfnamefont {A.}~\bibnamefont {Ran\c{c}on}}, \bibinfo
  {author} {\bibfnamefont {J.-F.}\ \bibnamefont {Cl\'ement}}, \bibinfo {author}
  {\bibfnamefont {P.}~\bibnamefont {Szriftgiser}}, \bibinfo {author}
  {\bibfnamefont {J.-C.}\ \bibnamefont {Garreau}}, \bibinfo {author}
  {\bibfnamefont {C.}~\bibnamefont {Tian}},\ and\ \bibinfo {author}
  {\bibfnamefont {R.}~\bibnamefont {Chicireanu}},\ }\bibfield  {title}
  {\bibinfo {title} {Experimental observation of a time-driven phase transition
  in quantum chaos},\ }\href {https://doi.org/10.1103/PhysRevLett.121.134101}
  {\bibfield  {journal} {\bibinfo  {journal} {Phys. Rev. Lett.}\ }\textbf
  {\bibinfo {volume} {121}},\ \bibinfo {pages} {134101} (\bibinfo {year}
  {2018}{\natexlab{c}})}\BibitemShut {NoStop}%
\bibitem [{\citenamefont {Grempel}\ \emph {et~al.}(1984)\citenamefont
  {Grempel}, \citenamefont {Prange},\ and\ \citenamefont
  {Fishman}}]{Fishman:LocDynAnderson:PRA84}%
  \BibitemOpen
  \bibfield  {author} {\bibinfo {author} {\bibfnamefont {D.~R.}\ \bibnamefont
  {Grempel}}, \bibinfo {author} {\bibfnamefont {R.~E.}\ \bibnamefont
  {Prange}},\ and\ \bibinfo {author} {\bibfnamefont {S.}~\bibnamefont
  {Fishman}},\ }\bibfield  {title} {\bibinfo {title} {{Quantum dynamics of a
  nonintegrable system}},\ }\href {https://doi.org/10.1103/PhysRevA.29.1639}
  {\bibfield  {journal} {\bibinfo  {journal} {Phys. Rev. A}\ }\textbf {\bibinfo
  {volume} {29}},\ \bibinfo {pages} {1639} (\bibinfo {year}
  {1984})}\BibitemShut {NoStop}%
\bibitem [{\citenamefont {Shepelyansky}(1986)}]{Shepelyansk:KRFloquet:PRL86}%
  \BibitemOpen
  \bibfield  {author} {\bibinfo {author} {\bibfnamefont {D.~L.}\ \bibnamefont
  {Shepelyansky}},\ }\bibfield  {title} {\bibinfo {title} {{Localization of
  quasienergy eigenfunctions in action space}},\ }\href
  {https://doi.org/10.1103/PhysRevLett.56.677} {\bibfield  {journal} {\bibinfo
  {journal} {Phys. Rev. Lett.}\ }\textbf {\bibinfo {volume} {56}},\ \bibinfo
  {pages} {677} (\bibinfo {year} {1986})}\BibitemShut {NoStop}%
\bibitem [{\citenamefont {Wigner}(1955)}]{Wigner1955}%
  \BibitemOpen
  \bibfield  {author} {\bibinfo {author} {\bibfnamefont {E.~P.}\ \bibnamefont
  {Wigner}},\ }\bibfield  {title} {\bibinfo {title} {Characteristic vectors of
  bordered matrices with infinite dimensions},\ }\href
  {http://www.jstor.org/stable/1970079} {\bibfield  {journal} {\bibinfo
  {journal} {Annals of Mathematics}\ }\textbf {\bibinfo {volume} {62}},\
  \bibinfo {pages} {548} (\bibinfo {year} {1955})}\BibitemShut {NoStop}%
\bibitem [{\citenamefont {Mehta}(2004)}]{MehtaBook2}%
  \BibitemOpen
  \bibfield  {author} {\bibinfo {author} {\bibfnamefont {M.~L.}\ \bibnamefont
  {Mehta}},\ }\href {https://doi.org//10.1016/C2009-0-22297-5} {\emph {\bibinfo
  {title} {Random matrices}}}\ (\bibinfo  {publisher} {Elsevier},\ \bibinfo
  {year} {2004})\BibitemShut {NoStop}%
\bibitem [{\citenamefont {Mirlin}(2000)}]{Mirlin2000}%
  \BibitemOpen
  \bibfield  {author} {\bibinfo {author} {\bibfnamefont {A.~D.}\ \bibnamefont
  {Mirlin}},\ }\bibfield  {title} {\bibinfo {title} {Statistics of energy
  levels and eigenfunctions in disordered systems},\ }\href
  {https://doi.org/https://doi.org/10.1016/S0370-1573(99)00091-5} {\bibfield
  {journal} {\bibinfo  {journal} {Physics Reports}\ }\textbf {\bibinfo {volume}
  {326}},\ \bibinfo {pages} {259 } (\bibinfo {year} {2000})}\BibitemShut
  {NoStop}%
\bibitem [{\citenamefont {Fyodorov}\ and\ \citenamefont
  {Mirlin}(1993{\natexlab{a}})}]{Fyodorov1993a}%
  \BibitemOpen
  \bibfield  {author} {\bibinfo {author} {\bibfnamefont {Y.}~\bibnamefont
  {Fyodorov}}\ and\ \bibinfo {author} {\bibfnamefont {A.}~\bibnamefont
  {Mirlin}},\ }\bibfield  {title} {\bibinfo {title} {Distribution of
  exponential decay rates of localized eigenfunctions in finite quasi-1d
  disordered systems},\ }\href@noop {} {\bibfield  {journal} {\bibinfo
  {journal} {Soviet Journal of Experimental and Theoretical Physics Letters}\
  }\textbf {\bibinfo {volume} {58}},\ \bibinfo {pages} {636} (\bibinfo {year}
  {1993}{\natexlab{a}})}\BibitemShut {NoStop}%
\bibitem [{\citenamefont {Casati}\ \emph {et~al.}(1990)\citenamefont {Casati},
  \citenamefont {Guarneri}, \citenamefont {Izrailev},\ and\ \citenamefont
  {Scharf}}]{Casati1991}%
  \BibitemOpen
  \bibfield  {author} {\bibinfo {author} {\bibfnamefont {G.}~\bibnamefont
  {Casati}}, \bibinfo {author} {\bibfnamefont {I.}~\bibnamefont {Guarneri}},
  \bibinfo {author} {\bibfnamefont {F.}~\bibnamefont {Izrailev}},\ and\
  \bibinfo {author} {\bibfnamefont {R.}~\bibnamefont {Scharf}},\ }\bibfield
  {title} {\bibinfo {title} {Scaling behavior of localization in quantum
  chaos},\ }\href {https://doi.org/10.1103/PhysRevLett.64.5} {\bibfield
  {journal} {\bibinfo  {journal} {Phys. Rev. Lett.}\ }\textbf {\bibinfo
  {volume} {64}},\ \bibinfo {pages} {5} (\bibinfo {year} {1990})}\BibitemShut
  {NoStop}%
\bibitem [{\citenamefont {Dittrich}\ and\ \citenamefont
  {Smilansky}(1991)}]{Dittrich1991}%
  \BibitemOpen
  \bibfield  {author} {\bibinfo {author} {\bibfnamefont {T.}~\bibnamefont
  {Dittrich}}\ and\ \bibinfo {author} {\bibfnamefont {U.}~\bibnamefont
  {Smilansky}},\ }\bibfield  {title} {\bibinfo {title} {Spectral properties of
  systems with dynamical localization. i. the local spectrum},\ }\href
  {https://doi.org/10.1088/0951-7715/4/1/006} {\bibfield  {journal} {\bibinfo
  {journal} {Nonlinearity}\ }\textbf {\bibinfo {volume} {4}},\ \bibinfo {pages}
  {59} (\bibinfo {year} {1991})}\BibitemShut {NoStop}%
\bibitem [{\citenamefont {Izrailev}(1995)}]{Izrailev1995}%
  \BibitemOpen
  \bibfield  {author} {\bibinfo {author} {\bibfnamefont {F.}~\bibnamefont
  {Izrailev}},\ }\bibfield  {title} {\bibinfo {title} {Scaling properties of
  spectra and eigenfunctions for quantum dynamical and disordered systems},\
  }\href {https://doi.org/https://doi.org/10.1016/0960-0779(94)E0063-U}
  {\bibfield  {journal} {\bibinfo  {journal} {Chaos, Solitons \& Fractals}\
  }\textbf {\bibinfo {volume} {5}},\ \bibinfo {pages} {1219} (\bibinfo {year}
  {1995})},\ \bibinfo {note} {quantum Chaos: Present and Future}\BibitemShut
  {NoStop}%
\bibitem [{\citenamefont {Shepelyansky}(1987)}]{Shepelyansky:Bicolor:PD87}%
  \BibitemOpen
  \bibfield  {author} {\bibinfo {author} {\bibfnamefont {D.~L.}\ \bibnamefont
  {Shepelyansky}},\ }\bibfield  {title} {\bibinfo {title} {{Localization of
  diffusive excitation in multi-level systems}},\ }\href
  {https://doi.org/http://dx.doi.org/10.1016/0167-2789(87)90123-0} {\bibfield
  {journal} {\bibinfo  {journal} {Physica D}\ }\textbf {\bibinfo {volume}
  {28}},\ \bibinfo {pages} {103} (\bibinfo {year} {1987})}\BibitemShut
  {NoStop}%
\bibitem [{\citenamefont {Rechester}\ \emph {et~al.}(1981)\citenamefont
  {Rechester}, \citenamefont {Rosenbluth},\ and\ \citenamefont
  {White}}]{Rechester:KRDiffCoeff:PRA81}%
  \BibitemOpen
  \bibfield  {author} {\bibinfo {author} {\bibfnamefont {A.~B.}\ \bibnamefont
  {Rechester}}, \bibinfo {author} {\bibfnamefont {M.~N.}\ \bibnamefont
  {Rosenbluth}},\ and\ \bibinfo {author} {\bibfnamefont {R.~B.}\ \bibnamefont
  {White}},\ }\bibfield  {title} {\bibinfo {title} {{Fourier-space paths
  applied to the calculation of diffusion for the Chirikov-Taylor model}},\
  }\href {https://doi.org/10.1103/PhysRevA.23.2664} {\bibfield  {journal}
  {\bibinfo  {journal} {Phys. Rev. A}\ }\textbf {\bibinfo {volume} {23}},\
  \bibinfo {pages} {2664} (\bibinfo {year} {1981})}\BibitemShut {NoStop}%
\bibitem [{\citenamefont {Rechester}\ and\ \citenamefont
  {White}(1980)}]{Rechester:Correl:PRL1980}%
  \BibitemOpen
  \bibfield  {author} {\bibinfo {author} {\bibfnamefont {A.~B.}\ \bibnamefont
  {Rechester}}\ and\ \bibinfo {author} {\bibfnamefont {R.~B.}\ \bibnamefont
  {White}},\ }\bibfield  {title} {\bibinfo {title} {Calculation of turbulent
  diffusion for the chirikov-taylor model},\ }\href
  {https://doi.org/10.1103/PhysRevLett.44.1586} {\bibfield  {journal} {\bibinfo
   {journal} {Phys. Rev. Lett.}\ }\textbf {\bibinfo {volume} {44}},\ \bibinfo
  {pages} {1586} (\bibinfo {year} {1980})}\BibitemShut {NoStop}%
\bibitem [{\citenamefont {Hainaut}\ \emph {et~al.}(2019)\citenamefont
  {Hainaut}, \citenamefont {Ran\c{c}on}, \citenamefont {Cl\'ement},
  \citenamefont {Manai}, \citenamefont {Szriftgiser}, \citenamefont {Delande},
  \citenamefont {Garreau},\ and\ \citenamefont
  {Chicireanu}}]{Hainaut2019IdealNJP}%
  \BibitemOpen
  \bibfield  {author} {\bibinfo {author} {\bibfnamefont {C.}~\bibnamefont
  {Hainaut}}, \bibinfo {author} {\bibfnamefont {A.}~\bibnamefont {Ran\c{c}on}},
  \bibinfo {author} {\bibfnamefont {J.-F.}\ \bibnamefont {Cl\'ement}}, \bibinfo
  {author} {\bibfnamefont {I.}~\bibnamefont {Manai}}, \bibinfo {author}
  {\bibfnamefont {P.}~\bibnamefont {Szriftgiser}}, \bibinfo {author}
  {\bibfnamefont {D.}~\bibnamefont {Delande}}, \bibinfo {author} {\bibfnamefont
  {J.~C.}\ \bibnamefont {Garreau}},\ and\ \bibinfo {author} {\bibfnamefont
  {R.}~\bibnamefont {Chicireanu}},\ }\bibfield  {title} {\bibinfo {title}
  {Experimental realization of an ideal floquet disordered system},\ }\href
  {https://doi.org/10.1088/1367-2630/ab0a79} {\bibfield  {journal} {\bibinfo
  {journal} {New Journal of Physics}\ }\textbf {\bibinfo {volume} {21}},\
  \bibinfo {pages} {035008} (\bibinfo {year} {2019})}\BibitemShut {NoStop}%
\bibitem [{\citenamefont {Fyodorov}\ and\ \citenamefont
  {Mirlin}(1993{\natexlab{b}})}]{Fyodorov1993}%
  \BibitemOpen
  \bibfield  {author} {\bibinfo {author} {\bibfnamefont {Y.~V.}\ \bibnamefont
  {Fyodorov}}\ and\ \bibinfo {author} {\bibfnamefont {A.~D.}\ \bibnamefont
  {Mirlin}},\ }\bibfield  {title} {\bibinfo {title} {Level-to-level
  fluctuations of the inverse participation ratio in finite quasi 1d disordered
  systems},\ }\href {https://doi.org/10.1103/PhysRevLett.71.412} {\bibfield
  {journal} {\bibinfo  {journal} {Phys. Rev. Lett.}\ }\textbf {\bibinfo
  {volume} {71}},\ \bibinfo {pages} {412} (\bibinfo {year}
  {1993}{\natexlab{b}})}\BibitemShut {NoStop}%
\bibitem [{\citenamefont {Pichard}(1991)}]{Pichard1991}%
  \BibitemOpen
  \bibfield  {author} {\bibinfo {author} {\bibfnamefont {J.-L.}\ \bibnamefont
  {Pichard}},\ }\bibinfo {title} {Random transfer matrix theory and conductance
  fluctuations},\ in\ \href {https://doi.org/10.1007/978-1-4899-3698-1_24}
  {\emph {\bibinfo {booktitle} {Quantum Coherence in Mesoscopic Systems}}},\
  \bibinfo {editor} {edited by\ \bibinfo {editor} {\bibfnamefont
  {B.}~\bibnamefont {Kramer}}}\ (\bibinfo  {publisher} {Springer US},\ \bibinfo
  {address} {Boston, MA},\ \bibinfo {year} {1991})\ pp.\ \bibinfo {pages}
  {369--400}\BibitemShut {NoStop}%
\bibitem [{\citenamefont {Lemari\'e}\ \emph {et~al.}(2009)\citenamefont
  {Lemari\'e}, \citenamefont {Chab\'e}, \citenamefont {Szriftgiser},
  \citenamefont {Garreau}, \citenamefont {Gr\'emaud},\ and\ \citenamefont
  {Delande}}]{Lemarie2009}%
  \BibitemOpen
  \bibfield  {author} {\bibinfo {author} {\bibfnamefont {G.}~\bibnamefont
  {Lemari\'e}}, \bibinfo {author} {\bibfnamefont {J.}~\bibnamefont {Chab\'e}},
  \bibinfo {author} {\bibfnamefont {P.}~\bibnamefont {Szriftgiser}}, \bibinfo
  {author} {\bibfnamefont {J.~C.}\ \bibnamefont {Garreau}}, \bibinfo {author}
  {\bibfnamefont {B.}~\bibnamefont {Gr\'emaud}},\ and\ \bibinfo {author}
  {\bibfnamefont {D.}~\bibnamefont {Delande}},\ }\bibfield  {title} {\bibinfo
  {title} {Observation of the anderson metal-insulator transition with atomic
  matter waves: Theory and experiment},\ }\href
  {https://doi.org/10.1103/PhysRevA.80.043626} {\bibfield  {journal} {\bibinfo
  {journal} {Phys. Rev. A}\ }\textbf {\bibinfo {volume} {80}},\ \bibinfo
  {pages} {043626} (\bibinfo {year} {2009})}\BibitemShut {NoStop}%
\bibitem [{\citenamefont {Tian}\ \emph {et~al.}(2005)\citenamefont {Tian},
  \citenamefont {Kamenev},\ and\ \citenamefont
  {Larkin}}]{Tian:EhrenfestTimeDynamicalLoc:PRB05}%
  \BibitemOpen
  \bibfield  {author} {\bibinfo {author} {\bibfnamefont {C.}~\bibnamefont
  {Tian}}, \bibinfo {author} {\bibfnamefont {A.}~\bibnamefont {Kamenev}},\ and\
  \bibinfo {author} {\bibfnamefont {A.}~\bibnamefont {Larkin}},\ }\bibfield
  {title} {\bibinfo {title} {{Ehrenfest time in the weak dynamical
  localization}},\ }\href {https://doi.org/10.1103/PhysRevB.72.045108}
  {\bibfield  {journal} {\bibinfo  {journal} {Phys. Rev. B}\ }\textbf {\bibinfo
  {volume} {72}},\ \bibinfo {pages} {045108} (\bibinfo {year}
  {2005})}\BibitemShut {NoStop}%
\bibitem [{\citenamefont {Sortais}\ \emph {et~al.}(2000)\citenamefont
  {Sortais}, \citenamefont {Bize}, \citenamefont {Nicolas}, \citenamefont
  {Clairon}, \citenamefont {Salomon},\ and\ \citenamefont
  {Williams}}]{Sortais2000}%
  \BibitemOpen
  \bibfield  {author} {\bibinfo {author} {\bibfnamefont {Y.}~\bibnamefont
  {Sortais}}, \bibinfo {author} {\bibfnamefont {S.}~\bibnamefont {Bize}},
  \bibinfo {author} {\bibfnamefont {C.}~\bibnamefont {Nicolas}}, \bibinfo
  {author} {\bibfnamefont {A.}~\bibnamefont {Clairon}}, \bibinfo {author}
  {\bibfnamefont {C.}~\bibnamefont {Salomon}},\ and\ \bibinfo {author}
  {\bibfnamefont {C.}~\bibnamefont {Williams}},\ }\bibfield  {title} {\bibinfo
  {title} {Cold collision frequency shifts in a ${}^{87}\mathrm{Rb}$ atomic
  fountain},\ }\href {https://doi.org/10.1103/PhysRevLett.85.3117} {\bibfield
  {journal} {\bibinfo  {journal} {Phys. Rev. Lett.}\ }\textbf {\bibinfo
  {volume} {85}},\ \bibinfo {pages} {3117} (\bibinfo {year}
  {2000})}\BibitemShut {NoStop}%
\end{thebibliography}%
\end{document}